\documentclass[10pt,journal,compsoc]{IEEEtran}

\newcommand{\xmark}{\ding{55}}
\usepackage{pifont}
\usepackage{amssymb}

\usepackage{cite}
\usepackage{enumitem}
\usepackage{soul}

\usepackage{cite}
\usepackage{soul,color}
\usepackage{hyperref}
\usepackage{makecell}
\usepackage{hhline}
\usepackage{ragged2e}
\usepackage{amsmath,varwidth,array,ragged2e}
\usepackage{multirow}
\usepackage{graphicx}
\usepackage{mathtools}  
\usepackage{mathtools}

\usepackage{amsmath}
\usepackage{amsthm}
\usepackage{amssymb}
\usepackage{array}

\usepackage{amsmath}
\usepackage{amsthm}
\usepackage{amssymb}
\usepackage[export]{adjustbox}
\usepackage{verbatim}
\usepackage{graphicx}
\usepackage{threeparttable}
\usepackage[table]{xcolor}
\usepackage{tikz}
\usetikzlibrary{shapes,arrows}
\tikzstyle{line}=[draw] 
\usepackage{amsfonts}
\usepackage{mathrsfs}
\usepackage[export]{adjustbox}
\usepackage{verbatim}
\usepackage{graphicx}

\UseRawInputEncoding

\usepackage[T1]{fontenc}
\usepackage{authblk}

\begin{document}
	%
	\title{A Comprehensive Review on Blockchains for Internet of Vehicles: Challenges and Directions}

\author{Brian Hildebrand, Mohamed Baza, 
Tara Salman, Fathi Amsaad, Abdul Razaqu and Abdullah Alourani

\thanks{ Brian Hildebrand is with GameAbove College of Engineering and Technology, Eastern Michigan University}

\thanks{Mohamed Baza is with the department of computer science at College of Charleston, Charleston, SC 29424}

\thanks{Tara Salman is with Department of Computer Science, Texas Tech University}

\thanks{Fathi Amsaad is with School of Information Security and Applied Computing (SISAC), Eastern Michigan University (EMU), MI 48197 USA}
}

	\markboth{IEEE TRANSACTIONS ON DEPENDAPLE AND SECURE COMPUTING,~Vol.~XX, No.~XX, November~2018}%
	{Shell \MakeLowercase{\textit{et al.}}: Bare Demo of IEEEtran.cls for Computer Society Journals}




\IEEEtitleabstractindextext{
    \begin{abstract}
    
    Internet of Vehicles (IoVs) consist of smart vehicles, Autonomous Vehicles (AVs) as well as roadside units (RSUs) that communicate wirelessly to provide enhanced transportation services such as improved traffic efficiency and reduced traffic congestion and accidents. IoVs, however, suffer from issues of security, privacy and trust.  Blockchain technology  has been emerged as  a decentralized approach for enhanced security without depending on trusted third parties to run services. Blockchain offers the benefits of trustworthiness, immutability, and mitigates the problem of single point of failure and other attacks. In this work, we present the state-of-the-art of \textbf{B}lockchain-enabled \textbf{IoV}s (\textbf{BIoV}) with a particular focus on their applications such as crowdsourcing-based applications, energy trading, traffic congestion reduction, collision and accident avoidance and infotainment and content cashing. We also present in-depth applications federated learning (FL) applications for BIoVs. The key challenges resulted from the integration of Blockchain with IoV is investigated in several domains such as edge computing, ML, and FL. Lastly,  a number of open issues and challenges as well as future opportunities in the area of AI-enabled BIoV, hardware-assisted security for BIoV and quantum computing attacks on BIoV.
    \end{abstract}
	

}

\maketitle
 
	\IEEEdisplaynontitleabstractindextext
	
	%
	\IEEEpeerreviewmaketitle


\section {Introduction} \label{sec:intro}

\IEEEPARstart{O}{ver}, the last two decades, Internet of Vehicles (IoV)  have been emerging as enabling cornerstone to the Intelligent Transportation Systems (ITSs)~\cite{shah2022blockchain}. Internet of Vehicles (IoV) comprise of smart vehicles and roadside units (RSUs) that communicate wirelessly to provide enhanced transportation services and capabilities such as traffic congestion reduction, accident avoidance and infotainment~\cite{xu2022accurate}.  This service are provided because vehicles contribute with data to an  traffic management center and with each other to build a spatio-temporal view of the traffic state~\cite{musa2022mobility,liu2022iov}.

Vehicles in IoV exchanges essential security messages to allow many novel applications such as crash avoidance, changing traffic routes, dynamic route selection, crisis cautioning,  traffic data, entertainment and content cashing, etc~\cite{lin2022joint,dong2022dependence,mageswari2022novel}. RSUs are set adjacent to the streets. They can provide street safety, routing, and administration to the other vehicles. However, traditional IoV follows  the traditional  centralized model. In this model, vehicles as well RSUs need to communicate directly with a central server to run the services mentioned earlier. This model is susceptible to single point of failure attacks. Also, raw data could be intercepted and modified by adversary during transmission. Moreover, the central server can be compromised by many cyberattacks including DOS attacks~\cite{gupta2022quantum,badr}.

Blockchain is a decentralized technology that has recently emerged to enable secure and trusted operations in cryptocurrency (Bitcoin) and financial applications. Blockchain has several beneficial properties such as immutability, transparency, and high efficiency which could be harnessed to enhance the security and preserve the privacy, integrity, and confidentiality of IoV applications \cite{zhang2022truthful,alrubei2022secure}. Due to its flexible and decentralized data structure, blockchain is not prone to a SPOF and is protected against centralized attacks. Other features of blockchain include immutability, transparency, and high efficiency.  Immutability allows the blockchain participating parties to establish mutual trust between communicating nodes in a decentralized blockchain network. Further, blockchain frameworks have evolved beyond cryptocurrencies to support smart contracts. Smart contracts are self-executing and self-enforcing programs that run on the blockchain without being administered by a trusted authority. Smart contracts provide a good avenue for implementing needed self-executing actions like managing blockchain updates for data exchange between vehicles and across the network \cite{lu2022stricts}.

The greater efficiency afforded by blockchain will also benefit IoVs since real-world transportation scenarios require decisions or actions to be made in real time. Failure in this could result in vehicular accidents or deaths. Therefore, this survey  aims to present the state of the art of IoVs that employ blockchain for various applications. We also discuss AI and machine learning applications for BIoVs. The Main contributions of this survey are summarized as follows.

\begin{itemize}

    \item  We provide a comprehensive survey of the techniques
for the integration of Blockchain and IoV 
to build a future intelligent transportation system, starting with a  background that includes the Blockchain technology, and IoV.

\item  We present the state-of-the-art research efforts and in depth discussion on the adoption of Blockchain for IoV scenarios, with the focus on applications such as energy trading, ride sharing, smart parking, vehicle platooning, accident avoidance and traffic congestion avoidance.

\item Blockchain-enabled IoV architectures are presented, including the  integration of Blockchain, edge computing,  automotive vehicles, and privacy preserving
approaches.

 \item The key challenges resulted from the integration of Blockchain  with IoV is investigated in several domains such as edge computing, and FL. 

\item Finally, the paper highlights a number of open issues and challenges as well as future  opportunities in the area of Blockchain, AI, Hardware security and Quantum Computing. 

\end{itemize}

\begin{table*}
    \centering
    \caption{Comparison of Blockchain-Based IoVs/IoV Surveys}
    \label{table:existing-BC-IoVs-surveys}
    \begin{tabular}{|l|c|c|c|c|c|c|c|c|c|c|c|c|c|c|}
    \hline
      & \multicolumn{10}{|c|}{Covered Topics} & 
      \makecell[c]{Future Opportunities \& Directions} & \makecell[c]{Covered Period of\\Reviewed Articles} \\
    \hline
     & \multicolumn{7}{|c|}{\makecell[c]{Blockchain-Based IoVs/IoV Applications}} & \multicolumn{3}{|c|}{Machine Learning} & & \\
    \hline
     &
     \rotatebox[origin=c]{90}{Energy Trading} &
     \rotatebox[origin=c]{90}{Ride Sharing} &
     \rotatebox[origin=c]{90}{Smart Parking} &
     \rotatebox[origin=c]{90}{Vehicle Platooning} &
     \rotatebox[origin=c]{90}{Infotainment} &
     \rotatebox[origin=c]{90}{Congestion Reduction} &
     \rotatebox[origin=c]{90}{Collision Avoidance} &
     \rotatebox[origin=c]{90}{AV} &
     \rotatebox[origin=c]{90}{Content Caching} &
     \rotatebox[origin=c]{90}{FL} & & \\
    \hline
    \cite{mollah1} & \checkmark & \checkmark & \checkmark & \checkmark & \checkmark & \checkmark & \xmark&\xmark &\xmark &\xmark & \makecell[l]{(1) Sharing road safety information\\(2) Content caching\\(3) Secure offload of block validation\\(4) Implementing PKI \& differential \\\hspace{4mm}privacy in Blockchain IoVs} & 2017-2020 \\
    \hline
    \cite{dibaei1} &\xmark &\xmark &\xmark & \checkmark & \checkmark & \checkmark & \checkmark & \checkmark & \xmark& \checkmark & \makecell[l]{(1) Integrate Blockchain with FL\\(2) Energy-saving consensus} & 2015-2020 \\
    \hline
    \cite{zuhair1} & \xmark& \xmark& \checkmark & \checkmark &\xmark &\xmark &\xmark &\xmark &\xmark & \xmark& \makecell[l]{(1) Increased throughput \& latency\\\hspace{4mm}introduced by Blockchain\\(2) Trust between vehicles\\(3) Cyber attacks\\(4) Scalability\\(5) Authentication\\(6) Connectivity} & 2018-2021 \\
    \hline
    \cite{karger1} & \checkmark & \checkmark & & \checkmark &\xmark & \xmark &\xmark &\xmark &\xmark& \xmark & \makecell[l]{(1) Consensus amongst vehicles\\(2) Detecting malicious vehicles\\(3) Privacy of participating vehicles\\(4) Trust between vehicles}& 2016-2021 \\
    \hline
    \cite{khoshavi1} & \checkmark & \checkmark & \checkmark & \checkmark & \xmark& \xmark& \checkmark & \checkmark & \xmark&\xmark & \makecell[l]{(1) Increased energy consumption,\\\hspace{4mm}resource use \& response time\\\hspace{4mm}introduced by Blockchain\\(2) Lightweight Blockchain}& 2016-2020\\

    \hline

    \makecell[l]{\cite{alladi1}} & \checkmark & \checkmark & \checkmark & \checkmark & \xmark & \xmark & \xmark & \checkmark & \xmark &  \xmark & \makecell[l]{(1) Scalability\\(2) Privacy\\(3) IoV prototyping \& simulation\\(4) Attacks on blockchain}& 2016-2021\\

    \hline
    \makecell[l]{Our\\Survey} & \checkmark & \checkmark & \checkmark & \checkmark & \checkmark & \checkmark & \checkmark & \checkmark & \checkmark & \checkmark & \makecell[l]{(1) Untraceable \& unexplainable\\\hspace{4mm}AI decisions\\(2) AI data \& model privacy\\(3) Hardware security related attacks\\(4) Quantum computing attacks} & 2008-2022\\
    \hline
    \end{tabular}
\end{table*}

\textbf{Related Surveys.} There exist several surveys that focus on Blockchain-based IoVs. Table \ref{fig:introBC} summarizes these surveys and the main differences between them. The work by \cite{mollah1} covers articles over a three year period beginning in 2017 and covers most of the Blockchain-based IoV applications such as energy trading, ride sharing, smart parking and vehicle platooning. The survey by Dibaei et al. \cite{dibaei1} focuses primarily on securing communication technologies in IoVs, including using ML and Blockchain defense mechanisms. Zuhair et al. \cite{zuhair1} provides an overview of architectures, characteristics and applications in IoV and provides a taxonomy for IoV attacks and possible countermeasures. The authors suggest that Blockchain can be used to enhance the security and privacy of IoV. However, they do not go too much into detail about Blockchain-based applications in IoV, providing reviews of a few articles on smart parking and vehicle platooning. Karger et al. \cite{karger1} provides a systematic literature review of Blockchain's role in mobility and transportation in smart cities.  The survey by Khoshavi et al. \cite{khoshavi1} focuses primarily on operation and security improvement aspects that Blockchain could provide for connected and autonomous vehicle (CAV) transportation systems. The authors discuss in detail Blockchain IoVs applications for energy trading, ride sharing, smart parking, vehicle platooning and collision avoidance. They also cover some aspects of AVs and ML. There also exist surveys on blockchain applications in smart cities \cite{xie1}, the energy sector \cite{bao2} and privacy management in social IoVs \cite{butt1}. Lastly, Baldini et al. \cite{baldini1} provides a survey of the use of distributed ledgers in transportation.

In summary, many of the preexisting surveys cover some of the Blockchain-based applications in IoV. Our survey provides detailed review to recent achievements on using Blockchain to secure IoV as well as focusing on incorporating ML and FL for Blockchain-based IoV. A related survey to us \cite{dibaei1} that discusses ML and FL applications for Blockchain-based IoVs but does not go depth  as our work and also our work provides recent published works till  2022 . To fill in the gaps in the research, our survey presents more details on Blockchain-based applications in IoV and discusses detailed Blockchain implementations of artificial intelligence (AI) or federated learning (FL) applications in IoV.

\textbf{Organization of Paper.} The rest of this paper is organized as follows. Section \ref{sec:background} provides an overview of Blockchain implementations for IoV, including an overview of relevant Blockchain consensus algorithms. Section \ref{sec:BC-based applications in IoV} presents an overview of Blockchain-based applications in IoV such as energy trading, ride sharing, vehicular cloud computing, and vehicular edge computing. Section \ref{sec:BC FL for IoVs} provides an overview of federated learning (FL) followed by Blockchain enhanced implementations of FL. Section \ref{sec:challenges future opportunities} outlines challenges and future directions for Blockchain-based IoVs and this is followed by a conclusion in section \ref{sec:conclusion}.

\section{Background} \label{sec:background}

In this section we provide an overview of IoV including the IoV architecture and IoV applications and the advantages of IoV. We also provide an overview of Blockchain technologies, including consensus algorithms relevant to IoVs.

\subsection{IoV Architecture and Overview}

\begin{figure}[ht]
    \centering
    \includegraphics[width=1\linewidth]{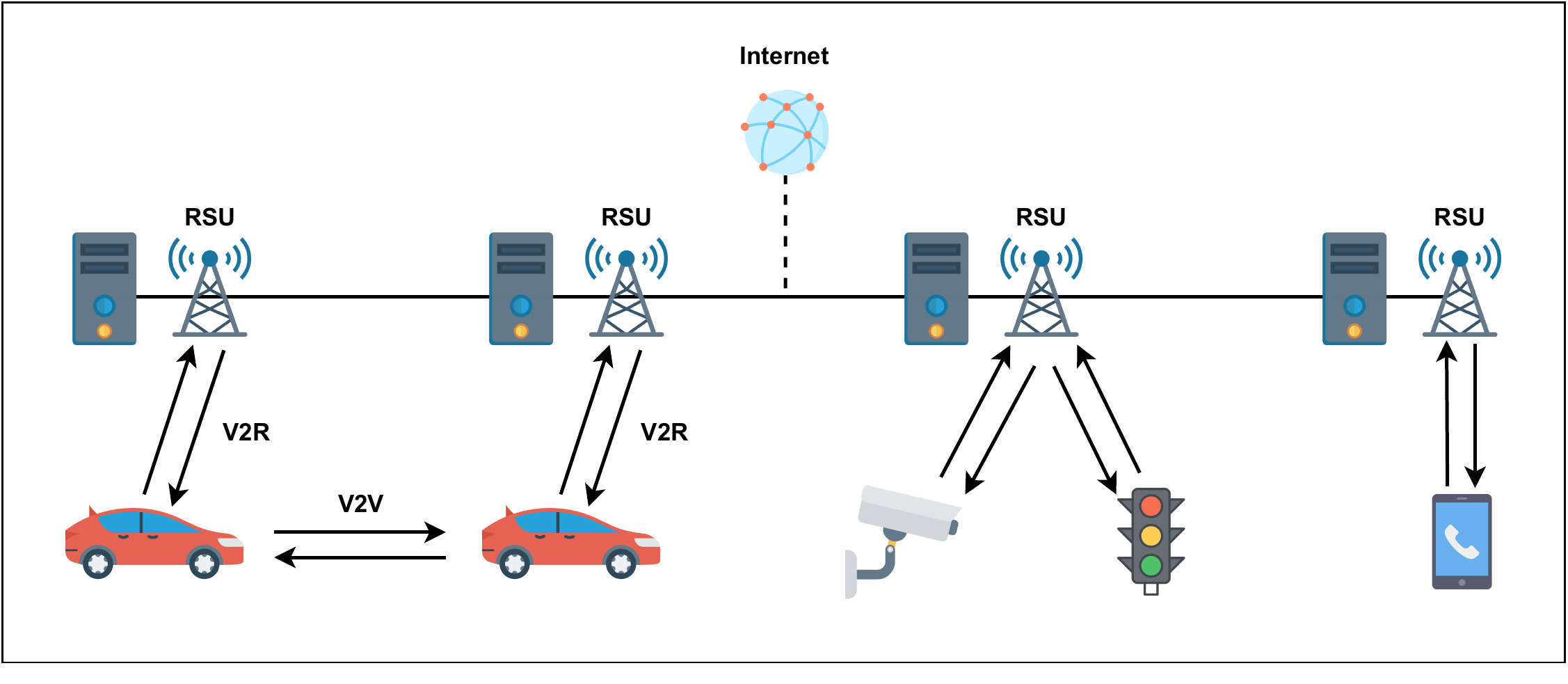}
    \caption{Typical IoV Infrastructure}
    \label{fig:iov}
\end{figure}

IoV is a vehicular ad hoc network (VANET) that has been connected to the Internet \cite{wan1}. VANETs are a type of mobile ad hoc network (MANET) comprised of smart vehicles and roadside units (RSUs) that communicate wirelessly, sharing information such as vehicle speeds and positions \cite{sakiz1}. IoVs can also be supplemented with AI and ML technologies to create an Intelligent Transportation System (ITS). Figure \ref{fig:iov} illustrates a typical IoV infrastructure where vehicles communicate with other vehicles using vehicle-to-vehicle (V2V) communication and communicate with RSUs using vehicle-to-RSU (V2R) communication. IoV is a vehicle-specific implementation of IoT that focuses primarily on connecting vehicles to the network infrastructure, other devices and the Internet. IoT refers to the networking of everyday objects such as home appliances equipped with sensors and computing capabilities with the goal of embedding technology into the background of everyday life.

By sharing important traffic information such as road conditions and traffic congestion between vehicles on the network, IoVs are capable of improving traffic efficiency and reducing traffic congestion and accidents \cite{yang1, liu1, mejri1, eze1}. In addition to safety-related applications, IoVs may also be designed to support other applications such as autonomous driving, infotainment, payment services, and vehicle usage tracking for calculating insurance rates \cite{lee1, sakiz1}. 

The primary advantages of IoV include improved traffic and fleet management, improved road safety, and better allocation of resources such as energy and fuel consumption. Implementing an IoV infrastructure would also increase the effectiveness and efficiency of autonomous driving applications. These advantages would also provide tangential benefits such as the reduced cost of goods and services that depend on our transportation system. For example, greater efficiency and better energy or fuel savings have the potential to lower the costs of food and products sold in stores since the price of such products takes into account delivery costs.

\subsection{Overview of Blockchain Technology}

\begin{figure}
    \centering
    \includegraphics[width=1\linewidth]{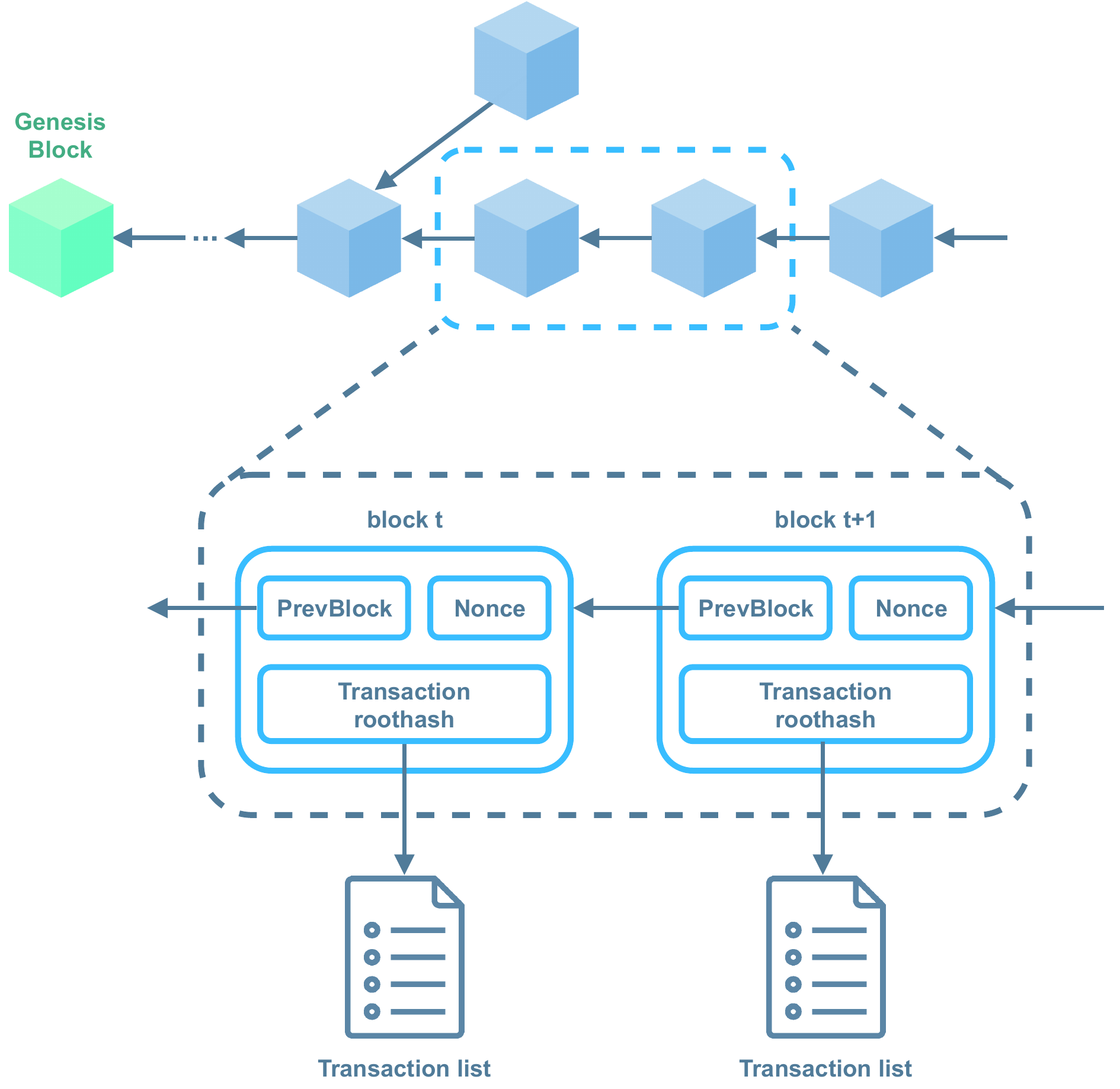}
    \caption{Illustration of a Blockchain structure, where transactions are packed into blocks that are linked to previous blocks.}
    \label{fig:introBC}
\end{figure}

A typical Blockchain system can be expressed as a collaboration of devices or nodes. However, the nodes do not necessarily trust one another. Together, the nodes maintain a set of shared global states and perform
transactions that may modify the states. Blockchain is a special type of data structure that is capable of storing historical states and transactions. All nodes in the system agree on the transactions and their order. Blockchain has been evolved to support   smart-contracts which are considered  an  autonomous computer  programs  that can run  on  a  Blockchain  network and its terms  can  be  pre-programmed with abilities such as self-enforcement without the interference from trusted authorities.

\subsubsection{Security Services of Blockchain}
Figure \ref{fig:introBC} depicts a Blockchain data structure. Each block in this structure is linked to its predecessor by way of a cryptographic pointer. The first block in the structure is called the genesis block. Due to this design, Blockchain is characterized as being a distributed ledger. Each block in Blockchain contains a roothash or Merkle root that contains the hashes of all transactions \cite{christidis2016Blockchains}. The transaction lifecycle is depicted in Fig.~\ref{fig:introBC} To add a block to Blockchain, a miner must successfully solve the computationally difficult puzzle of finding the right nonce 
for a block header. Blockchain offers several key features such as:

\begin{figure}
    \centering
    \includegraphics[width=0.8\linewidth]{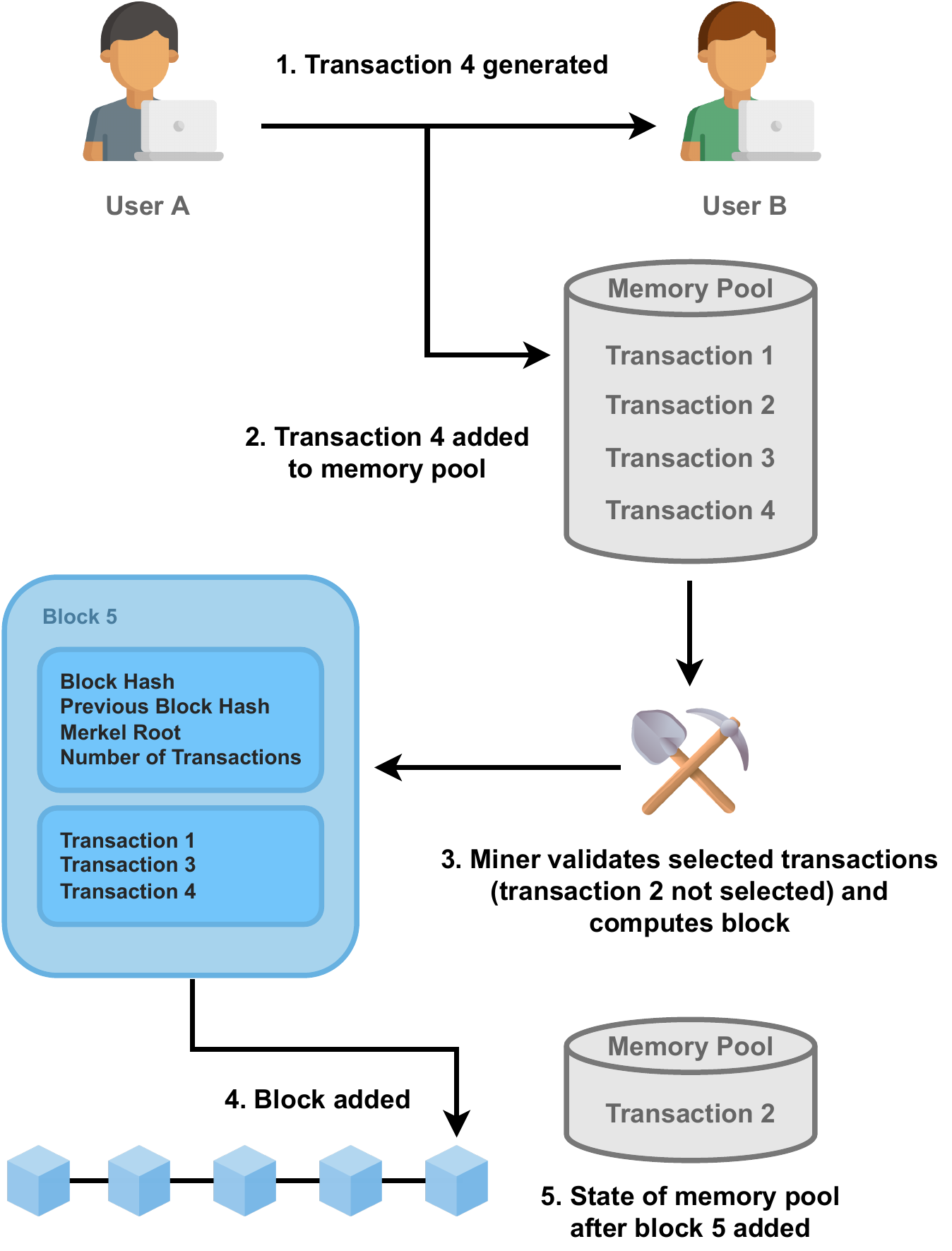}
    \caption{Illustration of a Blockchain Transaction Lifecycle. (1) A user A generates a transaction for user B. (2) Transaction is stored in memory pool along with other unconfirmed transactions. (3) A miner validates transactions from memory pool and computes block. Note that Transaction 2 was not valid so miners did not select it to add to block. (4) Valid block is added to Blockchain.}
    \label{fig:bc-transaction-lifecycle}
\end{figure}

\begin{itemize}

\item \textit{Decentralized data management.}  Every peer in the system has the authority to add data to the ledger, which is referred to as making a transaction. In this way, no one peer owns the Blockchain system more than another. IoVs would benefit from decentralization since it would remove obvious problems related to reliance on a central authority such as the potential for a single point of failure.

\item  \textit{Data security, tamper-proofness,
anti-forgeability and data integrity.} Blockchain is designed to store data in an immutable and tamper-proof fashion. The decentralized nature of Blockchain makes it is inordinately challenging for attackers to modify the ledger. Immutability would benefit IoVs in many ways such as providing immutable proof of vehicle transactions related to energy trading, ride sharing, smart parking and payment for leading a platoon.

\item  \textit{Transparency.}
A shared public list of transactions or exchanges of data provides every peer in the network access to every transaction ever made on the system. Therefore, the system is transparent. IoVs would particularly benefit from the transparency of exchanges in the context providing all parties and insurance companies transparent access to shared vehicular accident data.

\item  \textit{No central point of failure.}  The lack of a centralized storage system eliminates the possibility of losing data stored on the system by an attack of one node. This feature is particularly important in the context of IoV since failure of the system could result in accidents, loss of human life and economic losses.

\item  \textit{High efficiency.} Checking balances and completing transactions in a Blockchain system can, in theory, be instantaneous. High efficiency is particularly important in the context of IoV due to the dynamic and fast changing nature of vehicles on the road.

\end{itemize}

\subsubsection{Consensus Algorithms}

The content of Blockchain's distributed ledger consists of historical and current states. Any updates made to the ledger must be agreed upon by all parties. This agreement is referred to consensus. Several consensus algorithms exist for Blockchain. These algorithms can be characterized as computation-based or communication-based. Computation-based algorithms use proof of computation to randomly select a node and that node decides the next operation. In communication-based algorithms, each node has an equal vote and it takes multiple communication rounds to reach consensus. Bitcoin utilizes the Proof-of-Work (PoW) consensus algorithm, which is a purely computation-based. Practical Byzantine Fault Tolerance (PBFT) is an example of a purely communication-based consensus algorithm \cite{castro1}. Other consensus algorithms such as Proof of Elapsed Time (PoET), Proof of Authority (PoA) fall somewhere between computation-based and communication-based. Table \ref{table:consensus-algs-overview} provides an overview of the primary consensus algorithms relevant to IoVs.

In PoW \cite{nakamoto1}, the miner that expends the most amount of work or effort solving the hash puzzle is chosen to propose the new block. Therefore, in Blockchain systems that utilize PoW, miners that have greater computation power have a greater chance of being selected to propose a new block. Proof of Stake (PoS) was first introduced by Sunny King and Scott Nadel \cite{king1} in 2012 as a more scalable and energy efficient alternative to Bitcoin's PoW. In PoS, the ability to validate blocks is based on how many coins or network tokens the user has. PoS helps protect the Blockchain system from malicious users who wish to take over the majority of validation by requiring the malicious user to acquire a large amount of tokens in order to launch an attack on the system. Proof of Authority (PoA) \cite{wood1} is a modified version of PoS that uses the validator's identity in lieu of a user's stake. In PoA, transactions and blocks are validated by approved accounts known as validators. Under PoA, individuals must earn the right to become a validator based on their reputation. Individuals are motivated to support the transaction process to retain their reputation and status as a validator. Proof of Elapsed Time (PoET) \cite{intel1} was designed to be more efficient and use less resources and energy than conventional PoW. PoET also makes use of a fair lottery system to ensure efficient consensus process. The lottery system provides every participant a chance of being selected by spreading this chance equally amongst participants. Dan Larimer \cite{larimer1} first proposed Delegated Proof of Stake (DPoS) in 2014 as a faster and more energy efficient replacement for Bitcoin's PoW consensus algorithm. DPoS protects Blockchain better than PoW by employing an election process that ensures transactions receive better representation within the system. Users vote for delegates or witnesses who are responsible for generating and validating blocks. The primary difference between PoS and DPoS is that in PoS, users vote on the validity of blocks, while in DPoS, users elect delegates to validate the blocks.

Another notable Blockchain consensus algorithm is RAFT \cite{raft}, which stands for "Reliable, Replicated, Redundant And Fault-Tolerant." RAFT is based on the Paxos \cite{paxos1, paxos2} and was developed with the primary goal of increasing understandability and providing a better foundation for building practical systems. To make Blockchain consensus more understandable, RAFT separates the key elements of consensus such as leader selection, log replication and safety. RAFT also enforces a stronger degree of coherency in order to reduce the number of states that must be considered. In addition, RAFT affords enhanced safety using a mechanism for changing cluster membership to overlap majorities.

\textbf{Byzantine Fault Tolerant Consensus Algorithms.} Byzantine Fault Tolerance (BFT) refers to the ability of a system to resist failures resulting from the Byzantine Generals' Problem \cite{lamport1}. The Byzantine Generals' problem  is a logical dilemma first conceived of in 1982 where a number of generals must come to a consensus with the decision to retreat or attack. Communications between generals can be delayed, lost or destroyed or a general may act maliciously by sending fraudulent messages to bemuse other generals. With respect to Blockchain, generals refer to the different nodes or participants on Blockchain which may send false information or their communications may be delayed, lost or destroyed. A number of BFT-based algorithms have been proposed for Blockchain consensus such as Practical BFT (PBFT) \cite{castro1}, Redundant BFT (RBFT) \cite{aublin1}, Delegated BFT (dBFT) \cite{neo1} and BFT Delegated Proof of Stake (BFT-DPoS) \cite{eos1}.

 Castro and Liskov \cite{castro1} designed  Practical Byzantine Fault Tolerance (PBFT) for asynchronous systems to preserve liveness and safety while providing increased performance. Liveness means that clients receive replies to their requests while safety means that the replicated service satisfies linearizability \cite{herlihy1, castro2}. Concurrent operations are considered linearizable if they perform in a way that is equivalent to sequential operation. PBFT operates under the assumption that the maximum number of malicious nodes is no greater than or equal to one-third of all nodes on the system. Thus, as the number of nodes on the system increases, the system becomes more secure.  Aublin et al. \cite{aublin1} designed Redundant Byzantine Fault Tolerance (RBFT). RBFT achieves better performance during occasions where faults occur achieve this, RBFT executes multiple instances of BFT in parallel. RBFT is similar to PBFT but has an added procedure for transaction verification called "PROPAGATE".  
 
 Another consensus is called Byzantine Fault Tolerant Delegated Proof of Stake (BFT-DPoS) which is a combination of BFT and DPoS \cite{eos1}. Like DPoS, BFT-DPoS participants that hold tokens vote on block producers. The algorithm supplements traditional DPoS with BFT by allowing all producers to sign all blocks as long as no producer signs two blocks using the same timestamp or same block height. In addition to cryptocurrency, BFT-DPoS has been employed by authors such as \cite{fu1, fu2} for Blockchain-based vehicular networks, which are discussed in Section \ref{sec:collision-avoidance} below.

\begin{table*}[!t]
    \centering
    \caption{Overview of Consensus Algorithms used in Blockchain. Notice that public and permissionless Blockchains using PoW, PoS and DPoS have high scalability, low throughput and high confirmation times. In contrast, permissioned Blockchains using PBFT and RAFT have low scalability, low confirmation time and high throughput.}
    \label{table:consensus-algs-overview}
    \begin{tabular}{|l|c|c|c|c|c|}
    \hline
        \thead{Properties} &
        \thead{PoW} &
        \thead{PoS} &
        \thead{DPoS} &
        \thead{PBFT} &
        \thead{RAFT} \\
    \hhline{|=|=|=|=|=|=|}
    Blockchain Type & Permissionless & Permissionless & Permissionless & Permissioned & Permissioned \\
    \hline
    Participation Cost & Yes & Yes & Yes & No & No \\
    \hline
    Trust Model & Untrusted & Untrusted & Untrusted & Semi-trusted & Semi-trusted \\
    \hline
    Scalability & High & High & High & Low & Low \\
    \hline
    Throughput & $<$ 10 & $<$ 1,000 & $<$ 1,000 & $<$ 10,000 & $>$ 10,000 \\
    \hline
    Byzantine Fault Tolerance & 50\% & 50\% & 50\% & 33\% & - \\
    \hline
    Crash Fault Tolerance & 50\% & 50\% & 50\% & 33\% & 50\% \\
    \hline
    Confirmation Time & $>$ 100s & $<$ 100s & $<$ 100s & $<$ 10s & $<$ 10s \\
    \hline
    \end{tabular}
\end{table*}

Finally, Table \ref{table:consensus-algs-overview} provides a summary of the most prominent consensus algorithms employed in Blockchain. Overall, PoW provides good scalability but worse performance while BFT-based procotols such as PBFT offer good performance and limited scalability \cite{vukoli1}. PoS requires less energy than PoW and therefore is more befitting for large-scale Blockchain applications like IoT and IoV \cite{nguyen1}. Permissionless Blockchains that use PoW, PoS and DPoS offer high scalability but have low throughput and high confirmation times. On the other hand, permissioned Blockchains that use PBFT and RAFT offer little scalability but have high throughput and low confirmation time. The basic distinction between permissionless and permissioned Blockchains is that permissionless Blockchain allows anyone to participate in the Blockchain while permissioned Blockchain requires prior approval to participate.

\section{Blockchain-based IoV Applications}
\label{sec:BC-based applications in IoV}

\begin{figure*}[ht]
\centering

\tikzset{every picture/.style={line width=0.75pt}} 

\begin{tikzpicture}[x=0.545pt,y=0.545pt,yscale=-0.88,xscale=0.88]

\draw [color={rgb, 255:red, 172; green, 172; blue, 172 }  ,draw opacity=1 ]   (553.25,325.25) .. controls (593.25,295.25) and (545,163.25) .. (585,133.25) ;
\draw [color={rgb, 255:red, 172; green, 172; blue, 172 }  ,draw opacity=1 ]   (466.25,554.25) .. controls (503.5,646) and (406.5,716) .. (412.25,745.25) ;
\draw [color={rgb, 255:red, 172; green, 172; blue, 172 }  ,draw opacity=1 ]   (540.5,627) .. controls (526.5,719) and (434.5,733) .. (412.25,745.25) ;
\draw [color={rgb, 255:red, 172; green, 172; blue, 172 }  ,draw opacity=1 ]   (607.25,615.25) .. controls (646.5,639) and (625.5,649) .. (662.25,680.25) ;
\draw [color={rgb, 255:red, 172; green, 172; blue, 172 }  ,draw opacity=1 ]   (607.25,615.25) .. controls (646.5,639) and (652.75,625.75) .. (689.5,657) ;
\draw [color={rgb, 255:red, 172; green, 172; blue, 172 }  ,draw opacity=1 ]   (559.25,494.25) .. controls (600.5,515) and (663.5,626) .. (720.25,626.25) ;
\draw [color={rgb, 255:red, 172; green, 172; blue, 172 }  ,draw opacity=1 ]   (684.25,518) .. controls (707.5,581.75) and (730.5,589.25) .. (742.5,593.25) ;
\draw [color={rgb, 255:red, 172; green, 172; blue, 172 }  ,draw opacity=1 ]   (559.25,494.25) .. controls (634.5,529) and (748.5,507) .. (761.25,557.25) ;
\draw [color={rgb, 255:red, 172; green, 172; blue, 172 }  ,draw opacity=1 ]   (559.25,494.25) .. controls (598.5,518) and (570.5,584) .. (607.25,615.25) ;
\draw [color={rgb, 255:red, 172; green, 172; blue, 172 }  ,draw opacity=1 ]   (475.25,398.25) .. controls (478.5,437) and (514.5,496) .. (559.25,494.25) ;
\draw [color={rgb, 255:red, 172; green, 172; blue, 172 }  ,draw opacity=1 ]   (587.25,415.25) .. controls (669.5,409) and (706.5,350) .. (781.25,367) ;
\draw [color={rgb, 255:red, 172; green, 172; blue, 172 }  ,draw opacity=1 ]   (587.25,415.25) .. controls (648.5,414) and (729.5,447) .. (786.5,430) ;
\draw [color={rgb, 255:red, 172; green, 172; blue, 172 }  ,draw opacity=1 ]   (587.25,415.25) .. controls (636.5,413) and (700.5,509) .. (780.25,494.25) ;
\draw [color={rgb, 255:red, 172; green, 172; blue, 172 }  ,draw opacity=1 ]   (466.25,554.25) .. controls (455.5,639) and (284.5,654) .. (315.25,711.25) ;
\draw [color={rgb, 255:red, 172; green, 172; blue, 172 }  ,draw opacity=1 ]   (384.25,511.5) .. controls (361.5,571) and (308.5,644) .. (234.25,651.25) ;
\draw [color={rgb, 255:red, 172; green, 172; blue, 172 }  ,draw opacity=1 ]   (384.25,511.5) .. controls (350.5,560) and (247.5,632) .. (212.25,624.25) ;
\draw [color={rgb, 255:red, 172; green, 172; blue, 172 }  ,draw opacity=1 ]   (192.25,597.25) .. controls (199.5,555) and (379.5,543) .. (384.25,505.25) ;
\draw [color={rgb, 255:red, 172; green, 172; blue, 172 }  ,draw opacity=1 ]   (174.25,563.25) .. controls (168.5,520) and (295.5,527) .. (384.25,511.5) ;
\draw [color={rgb, 255:red, 172; green, 172; blue, 172 }  ,draw opacity=1 ]   (553.25,325.25) .. controls (593.25,295.25) and (718.25,326.25) .. (758.25,296.25) ;
\draw [color={rgb, 255:red, 172; green, 172; blue, 172 }  ,draw opacity=1 ]   (553.25,325.25) .. controls (593.25,295.25) and (701.75,299.25) .. (738.5,263.75) ;
\draw [color={rgb, 255:red, 172; green, 172; blue, 172 }  ,draw opacity=1 ]   (553.25,325.25) .. controls (593.25,295.25) and (690.5,272) .. (717.25,229.25) ;
\draw [color={rgb, 255:red, 172; green, 172; blue, 172 }  ,draw opacity=1 ]   (553.25,325.25) .. controls (593.25,295.25) and (646.25,223.25) .. (686.25,193.25) ;
\draw [color={rgb, 255:red, 172; green, 172; blue, 172 }  ,draw opacity=1 ]   (553.25,325.25) .. controls (593.25,295.25) and (597.25,188.25) .. (637.25,158.25) ;
\draw [color={rgb, 255:red, 172; green, 172; blue, 172 }  ,draw opacity=1 ]   (259.25,188.25) .. controls (296.5,207) and (351.5,324) .. (391.25,325.25) ;
\draw [color={rgb, 255:red, 172; green, 172; blue, 172 }  ,draw opacity=1 ]   (312.25,150.25) .. controls (298.5,199) and (407.5,269) .. (391.25,325.25) ;
\draw [color={rgb, 255:red, 172; green, 172; blue, 172 }  ,draw opacity=1 ]   (212.25,237.25) .. controls (255.5,229) and (338.5,356) .. (391.25,325.25) ;
\draw [color={rgb, 255:red, 172; green, 172; blue, 172 }  ,draw opacity=1 ]   (148.25,464.25) .. controls (222.5,492) and (301.5,413) .. (346.25,413.25) ;
\draw [color={rgb, 255:red, 172; green, 172; blue, 172 }  ,draw opacity=1 ]   (153.25,365.25) .. controls (221.5,348) and (268.5,423) .. (346.25,413.25) ;
\draw [color={rgb, 255:red, 172; green, 172; blue, 172 }  ,draw opacity=1 ]   (475.25,398.25) .. controls (491.5,461) and (546.5,366) .. (553.25,325.25) ;
\draw [color={rgb, 255:red, 172; green, 172; blue, 172 }  ,draw opacity=1 ]   (391.25,325.25) .. controls (397.5,353) and (461.5,456) .. (475.25,398.25) ;
\draw [color={rgb, 255:red, 172; green, 172; blue, 172 }  ,draw opacity=1 ]   (384.25,505.25) .. controls (419.5,511) and (480.5,451) .. (475.25,398.25) ;
\draw [color={rgb, 255:red, 172; green, 172; blue, 172 }  ,draw opacity=1 ]   (466.25,554.25) .. controls (453.5,536.25) and (475.5,466) .. (475.25,398.25) ;
\draw [color={rgb, 255:red, 172; green, 172; blue, 172 }  ,draw opacity=1 ]   (475.25,404.5) .. controls (478.5,443.25) and (539.5,432) .. (587.25,415.25) ;
\draw [color={rgb, 255:red, 172; green, 172; blue, 172 }  ,draw opacity=1 ]   (346.25,413.25) .. controls (386.25,383.25) and (460.5,458.75) .. (475.25,407.25) ;
\draw  [color={rgb, 255:red, 189; green, 188; blue, 188 }  ,draw opacity=1 ][fill={rgb, 255:red, 134; green, 240; blue, 126 }  ,fill opacity=1 ][line width=1.5]  (547,325.25) .. controls (547,321.8) and (549.8,319) .. (553.25,319) .. controls (556.7,319) and (559.5,321.8) .. (559.5,325.25) .. controls (559.5,328.7) and (556.7,331.5) .. (553.25,331.5) .. controls (549.8,331.5) and (547,328.7) .. (547,325.25) -- cycle ;
\draw  [color={rgb, 255:red, 189; green, 188; blue, 188 }  ,draw opacity=1 ][fill={rgb, 255:red, 240; green, 246; blue, 240 }  ,fill opacity=1 ][line width=1.5]  (469,398.25) .. controls (469,394.8) and (471.8,392) .. (475.25,392) .. controls (478.7,392) and (481.5,394.8) .. (481.5,398.25) .. controls (481.5,401.7) and (478.7,404.5) .. (475.25,404.5) .. controls (471.8,404.5) and (469,401.7) .. (469,398.25) -- cycle ;
\draw  [color={rgb, 255:red, 189; green, 188; blue, 188 }  ,draw opacity=1 ][fill={rgb, 255:red, 159; green, 191; blue, 241 }  ,fill opacity=1 ][line width=1.5]  (581,415.25) .. controls (581,411.8) and (583.8,409) .. (587.25,409) .. controls (590.7,409) and (593.5,411.8) .. (593.5,415.25) .. controls (593.5,418.7) and (590.7,421.5) .. (587.25,421.5) .. controls (583.8,421.5) and (581,418.7) .. (581,415.25) -- cycle ;
\draw  [color={rgb, 255:red, 189; green, 188; blue, 188 }  ,draw opacity=1 ][fill={rgb, 255:red, 219; green, 190; blue, 245 }  ,fill opacity=1 ][line width=1.5]  (378,511.5) .. controls (378,508.05) and (380.8,505.25) .. (384.25,505.25) .. controls (387.7,505.25) and (390.5,508.05) .. (390.5,511.5) .. controls (390.5,514.95) and (387.7,517.75) .. (384.25,517.75) .. controls (380.8,517.75) and (378,514.95) .. (378,511.5) -- cycle ;
\draw  [color={rgb, 255:red, 189; green, 188; blue, 188 }  ,draw opacity=1 ][fill={rgb, 255:red, 240; green, 192; blue, 126 }  ,fill opacity=1 ][line width=1.5]  (460,554.25) .. controls (460,550.8) and (462.8,548) .. (466.25,548) .. controls (469.7,548) and (472.5,550.8) .. (472.5,554.25) .. controls (472.5,557.7) and (469.7,560.5) .. (466.25,560.5) .. controls (462.8,560.5) and (460,557.7) .. (460,554.25) -- cycle ;
\draw  [color={rgb, 255:red, 189; green, 188; blue, 188 }  ,draw opacity=1 ][fill={rgb, 255:red, 129; green, 243; blue, 218 }  ,fill opacity=1 ][line width=1.5]  (340,413.25) .. controls (340,409.8) and (342.8,407) .. (346.25,407) .. controls (349.7,407) and (352.5,409.8) .. (352.5,413.25) .. controls (352.5,416.7) and (349.7,419.5) .. (346.25,419.5) .. controls (342.8,419.5) and (340,416.7) .. (340,413.25) -- cycle ;
\draw  [color={rgb, 255:red, 189; green, 188; blue, 188 }  ,draw opacity=1 ][fill={rgb, 255:red, 248; green, 231; blue, 28 }  ,fill opacity=1 ][line width=1.5]  (385,325.25) .. controls (385,321.8) and (387.8,319) .. (391.25,319) .. controls (394.7,319) and (397.5,321.8) .. (397.5,325.25) .. controls (397.5,328.7) and (394.7,331.5) .. (391.25,331.5) .. controls (387.8,331.5) and (385,328.7) .. (385,325.25) -- cycle ;
\draw  [color={rgb, 255:red, 189; green, 188; blue, 188 }  ,draw opacity=1 ][fill={rgb, 255:red, 134; green, 240; blue, 126 }  ,fill opacity=1 ][line width=1.5]  (631,158.25) .. controls (631,154.8) and (633.8,152) .. (637.25,152) .. controls (640.7,152) and (643.5,154.8) .. (643.5,158.25) .. controls (643.5,161.7) and (640.7,164.5) .. (637.25,164.5) .. controls (633.8,164.5) and (631,161.7) .. (631,158.25) -- cycle ;
\draw  [color={rgb, 255:red, 189; green, 188; blue, 188 }  ,draw opacity=1 ][fill={rgb, 255:red, 134; green, 240; blue, 126 }  ,fill opacity=1 ][line width=1.5]  (680,193.25) .. controls (680,189.8) and (682.8,187) .. (686.25,187) .. controls (689.7,187) and (692.5,189.8) .. (692.5,193.25) .. controls (692.5,196.7) and (689.7,199.5) .. (686.25,199.5) .. controls (682.8,199.5) and (680,196.7) .. (680,193.25) -- cycle ;
\draw  [color={rgb, 255:red, 189; green, 188; blue, 188 }  ,draw opacity=1 ][fill={rgb, 255:red, 134; green, 240; blue, 126 }  ,fill opacity=1 ][line width=1.5]  (711,229.25) .. controls (711,225.8) and (713.8,223) .. (717.25,223) .. controls (720.7,223) and (723.5,225.8) .. (723.5,229.25) .. controls (723.5,232.7) and (720.7,235.5) .. (717.25,235.5) .. controls (713.8,235.5) and (711,232.7) .. (711,229.25) -- cycle ;
\draw  [color={rgb, 255:red, 189; green, 188; blue, 188 }  ,draw opacity=1 ][fill={rgb, 255:red, 134; green, 240; blue, 126 }  ,fill opacity=1 ][line width=1.5]  (732.25,263.75) .. controls (732.25,260.3) and (735.05,257.5) .. (738.5,257.5) .. controls (741.95,257.5) and (744.75,260.3) .. (744.75,263.75) .. controls (744.75,267.2) and (741.95,270) .. (738.5,270) .. controls (735.05,270) and (732.25,267.2) .. (732.25,263.75) -- cycle ;
\draw  [color={rgb, 255:red, 189; green, 188; blue, 188 }  ,draw opacity=1 ][fill={rgb, 255:red, 134; green, 240; blue, 126 }  ,fill opacity=1 ][line width=1.5]  (752,296.25) .. controls (752,292.8) and (754.8,290) .. (758.25,290) .. controls (761.7,290) and (764.5,292.8) .. (764.5,296.25) .. controls (764.5,299.7) and (761.7,302.5) .. (758.25,302.5) .. controls (754.8,302.5) and (752,299.7) .. (752,296.25) -- cycle ;
\draw  [color={rgb, 255:red, 189; green, 188; blue, 188 }  ,draw opacity=1 ][fill={rgb, 255:red, 159; green, 191; blue, 241 }  ,fill opacity=1 ][line width=1.5]  (775,367) .. controls (775,363.55) and (777.8,360.75) .. (781.25,360.75) .. controls (784.7,360.75) and (787.5,363.55) .. (787.5,367) .. controls (787.5,370.45) and (784.7,373.25) .. (781.25,373.25) .. controls (777.8,373.25) and (775,370.45) .. (775,367) -- cycle ;
\draw  [color={rgb, 255:red, 189; green, 188; blue, 188 }  ,draw opacity=1 ][fill={rgb, 255:red, 159; green, 191; blue, 241 }  ,fill opacity=1 ][line width=1.5]  (780.25,430) .. controls (780.25,426.55) and (783.05,423.75) .. (786.5,423.75) .. controls (789.95,423.75) and (792.75,426.55) .. (792.75,430) .. controls (792.75,433.45) and (789.95,436.25) .. (786.5,436.25) .. controls (783.05,436.25) and (780.25,433.45) .. (780.25,430) -- cycle ;
\draw  [color={rgb, 255:red, 189; green, 188; blue, 188 }  ,draw opacity=1 ][fill={rgb, 255:red, 159; green, 191; blue, 241 }  ,fill opacity=1 ][line width=1.5]  (774,494.25) .. controls (774,490.8) and (776.8,488) .. (780.25,488) .. controls (783.7,488) and (786.5,490.8) .. (786.5,494.25) .. controls (786.5,497.7) and (783.7,500.5) .. (780.25,500.5) .. controls (776.8,500.5) and (774,497.7) .. (774,494.25) -- cycle ;
\draw  [color={rgb, 255:red, 189; green, 188; blue, 188 }  ,draw opacity=1 ][fill={rgb, 255:red, 219; green, 190; blue, 245 }  ,fill opacity=1 ][line width=1.5]  (186,597.25) .. controls (186,593.8) and (188.8,591) .. (192.25,591) .. controls (195.7,591) and (198.5,593.8) .. (198.5,597.25) .. controls (198.5,600.7) and (195.7,603.5) .. (192.25,603.5) .. controls (188.8,603.5) and (186,600.7) .. (186,597.25) -- cycle ;
\draw  [color={rgb, 255:red, 189; green, 188; blue, 188 }  ,draw opacity=1 ][fill={rgb, 255:red, 219; green, 190; blue, 245 }  ,fill opacity=1 ][line width=1.5]  (206,624.25) .. controls (206,620.8) and (208.8,618) .. (212.25,618) .. controls (215.7,618) and (218.5,620.8) .. (218.5,624.25) .. controls (218.5,627.7) and (215.7,630.5) .. (212.25,630.5) .. controls (208.8,630.5) and (206,627.7) .. (206,624.25) -- cycle ;
\draw  [color={rgb, 255:red, 189; green, 188; blue, 188 }  ,draw opacity=1 ][fill={rgb, 255:red, 219; green, 190; blue, 245 }  ,fill opacity=1 ][line width=1.5]  (228,651.25) .. controls (228,647.8) and (230.8,645) .. (234.25,645) .. controls (237.7,645) and (240.5,647.8) .. (240.5,651.25) .. controls (240.5,654.7) and (237.7,657.5) .. (234.25,657.5) .. controls (230.8,657.5) and (228,654.7) .. (228,651.25) -- cycle ;
\draw  [color={rgb, 255:red, 189; green, 188; blue, 188 }  ,draw opacity=1 ][fill={rgb, 255:red, 219; green, 190; blue, 245 }  ,fill opacity=1 ][line width=1.5]  (168,563.25) .. controls (168,559.8) and (170.8,557) .. (174.25,557) .. controls (177.7,557) and (180.5,559.8) .. (180.5,563.25) .. controls (180.5,566.7) and (177.7,569.5) .. (174.25,569.5) .. controls (170.8,569.5) and (168,566.7) .. (168,563.25) -- cycle ;
\draw  [color={rgb, 255:red, 189; green, 188; blue, 188 }  ,draw opacity=1 ][fill={rgb, 255:red, 238; green, 122; blue, 164 }  ,fill opacity=1 ][line width=1.5]  (406,745.25) .. controls (406,741.8) and (408.8,739) .. (412.25,739) .. controls (415.7,739) and (418.5,741.8) .. (418.5,745.25) .. controls (418.5,748.7) and (415.7,751.5) .. (412.25,751.5) .. controls (408.8,751.5) and (406,748.7) .. (406,745.25) -- cycle ;
\draw  [color={rgb, 255:red, 189; green, 188; blue, 188 }  ,draw opacity=1 ][fill={rgb, 255:red, 248; green, 231; blue, 28 }  ,fill opacity=1 ][line width=1.5]  (253,188.25) .. controls (253,184.8) and (255.8,182) .. (259.25,182) .. controls (262.7,182) and (265.5,184.8) .. (265.5,188.25) .. controls (265.5,191.7) and (262.7,194.5) .. (259.25,194.5) .. controls (255.8,194.5) and (253,191.7) .. (253,188.25) -- cycle ;
\draw  [color={rgb, 255:red, 189; green, 188; blue, 188 }  ,draw opacity=1 ][fill={rgb, 255:red, 248; green, 231; blue, 28 }  ,fill opacity=1 ][line width=1.5]  (306,150.25) .. controls (306,146.8) and (308.8,144) .. (312.25,144) .. controls (315.7,144) and (318.5,146.8) .. (318.5,150.25) .. controls (318.5,153.7) and (315.7,156.5) .. (312.25,156.5) .. controls (308.8,156.5) and (306,153.7) .. (306,150.25) -- cycle ;
\draw  [color={rgb, 255:red, 189; green, 188; blue, 188 }  ,draw opacity=1 ][fill={rgb, 255:red, 248; green, 231; blue, 28 }  ,fill opacity=1 ][line width=1.5]  (206,237.25) .. controls (206,233.8) and (208.8,231) .. (212.25,231) .. controls (215.7,231) and (218.5,233.8) .. (218.5,237.25) .. controls (218.5,240.7) and (215.7,243.5) .. (212.25,243.5) .. controls (208.8,243.5) and (206,240.7) .. (206,237.25) -- cycle ;
\draw  [color={rgb, 255:red, 189; green, 188; blue, 188 }  ,draw opacity=1 ][fill={rgb, 255:red, 129; green, 243; blue, 218 }  ,fill opacity=1 ][line width=1.5]  (147,365.25) .. controls (147,361.8) and (149.8,359) .. (153.25,359) .. controls (156.7,359) and (159.5,361.8) .. (159.5,365.25) .. controls (159.5,368.7) and (156.7,371.5) .. (153.25,371.5) .. controls (149.8,371.5) and (147,368.7) .. (147,365.25) -- cycle ;
\draw  [color={rgb, 255:red, 189; green, 188; blue, 188 }  ,draw opacity=1 ][fill={rgb, 255:red, 129; green, 243; blue, 218 }  ,fill opacity=1 ][line width=1.5]  (142,464.25) .. controls (142,460.8) and (144.8,458) .. (148.25,458) .. controls (151.7,458) and (154.5,460.8) .. (154.5,464.25) .. controls (154.5,467.7) and (151.7,470.5) .. (148.25,470.5) .. controls (144.8,470.5) and (142,467.7) .. (142,464.25) -- cycle ;
\draw  [color={rgb, 255:red, 189; green, 188; blue, 188 }  ,draw opacity=1 ][fill={rgb, 255:red, 240; green, 192; blue, 126 }  ,fill opacity=1 ][line width=1.5]  (309,711.25) .. controls (309,707.8) and (311.8,705) .. (315.25,705) .. controls (318.7,705) and (321.5,707.8) .. (321.5,711.25) .. controls (321.5,714.7) and (318.7,717.5) .. (315.25,717.5) .. controls (311.8,717.5) and (309,714.7) .. (309,711.25) -- cycle ;
\draw  [color={rgb, 255:red, 189; green, 188; blue, 188 }  ,draw opacity=1 ][fill={rgb, 255:red, 238; green, 122; blue, 164 }  ,fill opacity=1 ][line width=1.5]  (553,494.25) .. controls (553,490.8) and (555.8,488) .. (559.25,488) .. controls (562.7,488) and (565.5,490.8) .. (565.5,494.25) .. controls (565.5,497.7) and (562.7,500.5) .. (559.25,500.5) .. controls (555.8,500.5) and (553,497.7) .. (553,494.25) -- cycle ;
\draw  [color={rgb, 255:red, 189; green, 188; blue, 188 }  ,draw opacity=1 ][fill={rgb, 255:red, 238; green, 122; blue, 164 }  ,fill opacity=1 ][line width=1.5]  (755,557.25) .. controls (755,553.8) and (757.8,551) .. (761.25,551) .. controls (764.7,551) and (767.5,553.8) .. (767.5,557.25) .. controls (767.5,560.7) and (764.7,563.5) .. (761.25,563.5) .. controls (757.8,563.5) and (755,560.7) .. (755,557.25) -- cycle ;
\draw  [color={rgb, 255:red, 189; green, 188; blue, 188 }  ,draw opacity=1 ][fill={rgb, 255:red, 238; green, 122; blue, 164 }  ,fill opacity=1 ][line width=1.5]  (714,626.25) .. controls (714,622.8) and (716.8,620) .. (720.25,620) .. controls (723.7,620) and (726.5,622.8) .. (726.5,626.25) .. controls (726.5,629.7) and (723.7,632.5) .. (720.25,632.5) .. controls (716.8,632.5) and (714,629.7) .. (714,626.25) -- cycle ;
\draw  [color={rgb, 255:red, 189; green, 188; blue, 188 }  ,draw opacity=1 ][fill={rgb, 255:red, 238; green, 122; blue, 164 }  ,fill opacity=1 ][line width=1.5]  (601,615.25) .. controls (601,611.8) and (603.8,609) .. (607.25,609) .. controls (610.7,609) and (613.5,611.8) .. (613.5,615.25) .. controls (613.5,618.7) and (610.7,621.5) .. (607.25,621.5) .. controls (603.8,621.5) and (601,618.7) .. (601,615.25) -- cycle ;
\draw  [color={rgb, 255:red, 189; green, 188; blue, 188 }  ,draw opacity=1 ][fill={rgb, 255:red, 238; green, 122; blue, 164 }  ,fill opacity=1 ][line width=1.5]  (678,520.25) .. controls (678,516.8) and (680.8,514) .. (684.25,514) .. controls (687.7,514) and (690.5,516.8) .. (690.5,520.25) .. controls (690.5,523.7) and (687.7,526.5) .. (684.25,526.5) .. controls (680.8,526.5) and (678,523.7) .. (678,520.25) -- cycle ;
\draw  [color={rgb, 255:red, 189; green, 188; blue, 188 }  ,draw opacity=1 ][fill={rgb, 255:red, 238; green, 122; blue, 164 }  ,fill opacity=1 ][line width=1.5]  (736.25,593.25) .. controls (736.25,589.8) and (739.05,587) .. (742.5,587) .. controls (745.95,587) and (748.75,589.8) .. (748.75,593.25) .. controls (748.75,596.7) and (745.95,599.5) .. (742.5,599.5) .. controls (739.05,599.5) and (736.25,596.7) .. (736.25,593.25) -- cycle ;
\draw  [color={rgb, 255:red, 189; green, 188; blue, 188 }  ,draw opacity=1 ][fill={rgb, 255:red, 238; green, 122; blue, 164 }  ,fill opacity=1 ][line width=1.5]  (644,585.25) .. controls (644,581.8) and (646.8,579) .. (650.25,579) .. controls (653.7,579) and (656.5,581.8) .. (656.5,585.25) .. controls (656.5,588.7) and (653.7,591.5) .. (650.25,591.5) .. controls (646.8,591.5) and (644,588.7) .. (644,585.25) -- cycle ;
\draw  [color={rgb, 255:red, 189; green, 188; blue, 188 }  ,draw opacity=1 ][fill={rgb, 255:red, 238; green, 122; blue, 164 }  ,fill opacity=1 ][line width=1.5]  (683.25,657) .. controls (683.25,653.55) and (686.05,650.75) .. (689.5,650.75) .. controls (692.95,650.75) and (695.75,653.55) .. (695.75,657) .. controls (695.75,660.45) and (692.95,663.25) .. (689.5,663.25) .. controls (686.05,663.25) and (683.25,660.45) .. (683.25,657) -- cycle ;
\draw  [color={rgb, 255:red, 189; green, 188; blue, 188 }  ,draw opacity=1 ][fill={rgb, 255:red, 238; green, 122; blue, 164 }  ,fill opacity=1 ][line width=1.5]  (656,680.25) .. controls (656,676.8) and (658.8,674) .. (662.25,674) .. controls (665.7,674) and (668.5,676.8) .. (668.5,680.25) .. controls (668.5,683.7) and (665.7,686.5) .. (662.25,686.5) .. controls (658.8,686.5) and (656,683.7) .. (656,680.25) -- cycle ;
\draw [color={rgb, 255:red, 172; green, 172; blue, 172 }  ,draw opacity=1 ]   (559.25,494.25) .. controls (598.5,518) and (533.5,591) .. (540.5,627) ;
\draw  [color={rgb, 255:red, 189; green, 188; blue, 188 }  ,draw opacity=1 ][fill={rgb, 255:red, 238; green, 122; blue, 164 }  ,fill opacity=1 ][line width=1.5]  (534.25,627) .. controls (534.25,623.55) and (537.05,620.75) .. (540.5,620.75) .. controls (543.95,620.75) and (546.75,623.55) .. (546.75,627) .. controls (546.75,630.45) and (543.95,633.25) .. (540.5,633.25) .. controls (537.05,633.25) and (534.25,630.45) .. (534.25,627) -- cycle ;
\draw  [color={rgb, 255:red, 189; green, 188; blue, 188 }  ,draw opacity=1 ][fill={rgb, 255:red, 134; green, 240; blue, 126 }  ,fill opacity=1 ][line width=1.5]  (578.75,133.25) .. controls (578.75,129.8) and (581.55,127) .. (585,127) .. controls (588.45,127) and (591.25,129.8) .. (591.25,133.25) .. controls (591.25,136.7) and (588.45,139.5) .. (585,139.5) .. controls (581.55,139.5) and (578.75,136.7) .. (578.75,133.25) -- cycle ;

\draw (530,285) node [anchor=north west][inner sep=0.75pt]  [rotate=-318.01] [align=left] {Energy Trading};
\draw (652,400) node [anchor=north west][inner sep=0.75pt]   [align=left] {Crowd Sourcing};
\draw (240,570) node [anchor=north west][inner sep=0.75pt]  [rotate=-323.2] [align=left] {Vehicle Platooning};
\draw (387,680) node [anchor=north west][inner sep=0.75pt]  [rotate=-307.95] [align=left] {Infotainment};
\draw (280,180) node [anchor=north west][inner sep=0.75pt]  [rotate=-48.01] [align=left] {\begin{minipage}[lt]{84.94pt}\setlength\topsep{0pt}
\begin{center}
Traffic Congestion\\Reduction
\end{center}

\end{minipage}};
\draw (170,380) node [anchor=north west][inner sep=0.75pt]   [align=left] {\begin{minipage}[lt]{92.3pt}\setlength\topsep{0pt}
\begin{center}
Collision \& Accident\\Avoidance
\end{center}

\end{minipage}};
\draw (628.44,127.89) node [anchor=north west][inner sep=0.75pt]  [rotate=-311.5] [align=left] {\href{}{\cite{aggarwal1} Aggarwal (2020)}\\\cite{long1} Long (2020)\\\cite{hassija1} Hassija (2020)};
\draw (772.68,282.26) node [anchor=north west][inner sep=0.75pt]  [rotate=-341.78] [align=left] {\cite{pustisek1} Pustisek (2016)};
\draw (715.59,207.24) node [anchor=north west][inner sep=0.75pt]  [rotate=-323.16] [align=left] {\cite{knirsch1} Knirsch (2018)\\\cite{su1} Su (2018)};

\draw (752.09,249.1) node [anchor=north west][inner sep=0.75pt]  [rotate=-331.75] [align=left] {\cite{kang1} Kang (2017)};
\draw (691.57,174.61) node [anchor=north west][inner sep=0.75pt]  [rotate=-318.98] [align=left] {\cite{huang1} Huang (2019)};

\draw (794.46,344.71) node [anchor=north west][inner sep=0.75pt]  [rotate=-353.07] [align=left] {\cite{li3} Li (2020)\\\cite{wang1} Wang (2020)};
\draw (800.9,399.38) node [anchor=north west][inner sep=0.75pt]   [align=left] {\cite{baza2019b} Baza (2019)\\\cite{li5} Li (2019)\\\cite{aljaroodi1} Al-Jaroodi (2019)};
\draw (799.12,490.34) node [anchor=north west][inner sep=0.75pt]  [rotate=-11.71] [align=left] {\cite{badr} Badr (2021)};
\draw (9,647) node [anchor=north west][inner sep=0.75pt]  [rotate=-330.07] [align=left] {\cite{wagner1} Wagner (2018)};
\draw (55,671) node [anchor=north west][inner sep=0.75pt]  [rotate=-326.81] [align=left] {\cite{ying2} Ying (2019)};
\draw (75,710) node [anchor=north west][inner sep=0.75pt]  [rotate=-325.38] [align=left] {\cite{chen1} Chen (2020)};
\draw (125,732) node [anchor=north west][inner sep=0.75pt]  [rotate=-319.73] [align=left] {\cite{li10} Li (2021)};
\draw (210,825) node [anchor=north west][inner sep=0.75pt]  [rotate=-309.18] [align=left] {\cite{kim2} Kim (2021)};
\draw (245,34) node [anchor=north west][inner sep=0.75pt]  [rotate=-55.64] [align=left] {\cite{li9} Li (2018)};
\draw (160,75) node [anchor=north west][inner sep=0.75pt]  [rotate=-45.23] [align=left] {\cite{liu1} Liu (2019)};
\draw (90,130) node [anchor=north west][inner sep=0.75pt]  [rotate=-36.48] [align=left] {\cite{yang1} Yang (2020)};

\draw (-66,319) node [anchor=north west][inner sep=0.75pt]  [rotate=-2.96] [align=left] {\begin{minipage}[lt]{100pt}\setlength\topsep{0pt}
\begin{flushright}
\cite{westerlund1} Westerlund (2020)\\\cite{singh4} Singh (2020)\\\cite{fu2} Fu (2020)
\end{flushright}

\end{minipage}};
\draw (-45,451) node [anchor=north west][inner sep=0.75pt]  [rotate=-357.34] [align=left] {\cite{8969653} Buzachis (2018)\\\cite{8969653} Buzachis (2019)};
\draw (350,910) node [anchor=north west][inner sep=0.75pt]  [rotate=-283.73] [align=left] {\begin{minipage}[lt]{76.59pt}\setlength\topsep{0pt}
\begin{flushright}
\cite{dai1} Dai (2020)\\\cite{qian1} Qian (2020)
\end{flushright}

\end{minipage}};
\draw (758.45,587.69) node [anchor=north west][inner sep=0.75pt]  [rotate=-29.26] [align=left] {\cite{zhang1} Zhang (2019)};
\draw (570,450) node [anchor=north west][inner sep=0.75pt]  [rotate=-42.81] [align=left] {Machine Learning};
\draw (683,482) node [anchor=north west][inner sep=0.75pt]  [rotate=-24] [align=left] {{\small Q-Learning}};
\draw (671.99,576.36) node [anchor=north west][inner sep=0.75pt]  [rotate=-39.21] [align=left] {{\small SVM}};
\draw (738.28,625.72) node [anchor=north west][inner sep=0.75pt]  [rotate=-33.54] [align=left] {\cite{shen1} Shen (2020)};
\draw (620,583) node [anchor=north west][inner sep=0.75pt]  [rotate=-49.58] [align=left] {{\small AV}};
\draw (778.83,554.16) node [anchor=north west][inner sep=0.75pt]  [rotate=-23.92] [align=left] {\cite{qui1} Qui (2018)};
\draw (712.13,652.18) node [anchor=north west][inner sep=0.75pt]  [rotate=-41.05] [align=left] {\cite{fu2} Fu (2020)\\\cite{pokhrel2} Pokhrel (2020)};
\draw (678.9,687.3) node [anchor=north west][inner sep=0.75pt]  [rotate=-52.52] [align=left] {\cite{kumar1} Kumar (2021)};
\draw (563,600) node [anchor=north west][inner sep=0.75pt]  [rotate=-59.76] [align=left] {{\small content caching}};
\draw (578.69,106.58) node [anchor=north west][inner sep=0.75pt]  [rotate=-304.64] [align=left] {\cite{wu1} Wu (2022)\\\cite{li11} Li (2022)};

\end{tikzpicture}
\caption{Applications of Blockchain-Enabled IoV}
\label{fig:bc-iov-apps}
\end{figure*}
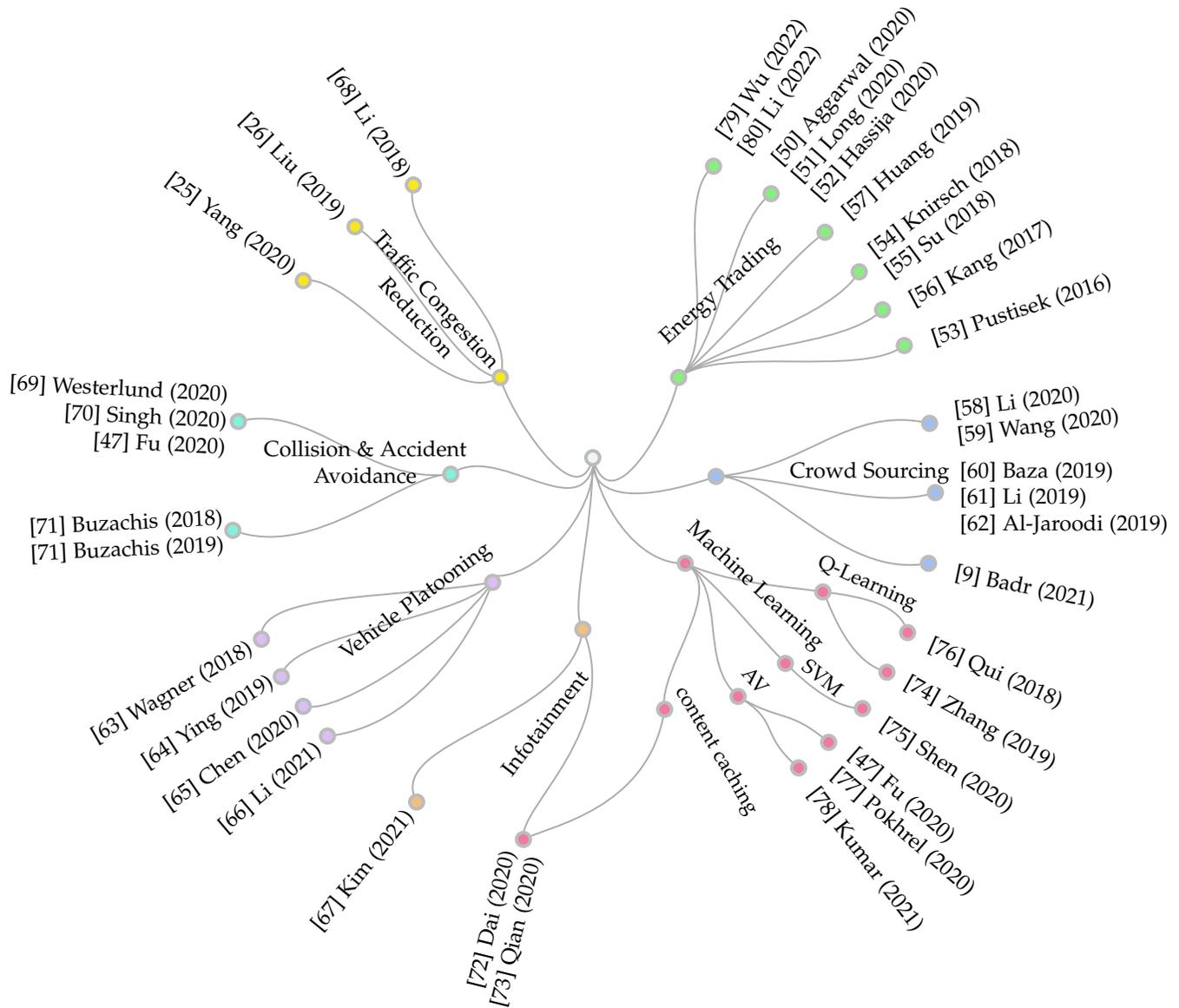

In this section, we review the primary applications of Blockchain-based IoVs. Figure \ref{fig:bc-iov-apps} summarizes  the current achievements in BIoV  applications.

\subsection{Energy Trading}

With the advancement and increase in the number of electric vehicles (EVs), vehicle-to-grid (V2G) networks have been developed that are capable of providing a two-way flow of electrical energy between EVs and charging stations within smart grids \cite{aggarwal1}. Energy trading is the mechanism used by V2Gs that permits charging and discharging of EVs and other entities and serves to manage energy demand \cite{kaur1}. In an V2G network, EVs can act as either energy consumers by charging their batteries at charging stations or energy producers by discharging surplus electricity back into the system. In this way, energy trading in V2G networks can serve to enhance overall energy efficiency and greater sustainability \cite{long1, wang6}.

Traditional V2Gs, however, rely on a central server, which is susceptible to SPOF. Therefore, a number of researchers have proposed implementing V2Gs using Blockchain \cite{aggarwal1, long1, hassija1}. In addition to solving the centralization problem, Blockchain can be utilized to enhance security and privacy of EVs as well as offering the Blockchain benefits of decentralization, security, immutability and transparency. However, there are several  Challenges to integrate Blockchain in V2G networks for energy trading (See Table \ref{table:energy-trading-BC-challenges}) that include securing the identity and location privacy of EVs during energy trading while at the same time ensuring the efficient distribution of energy amongst V2G entities. Another challenge to integrate Blockchain in V2G networks used for energy trading is ensuring the scalability of the V2G network to support the dynamic nature of EVs over the network, including the increasing number of transactions on the network. 

\begin{table}[!t]
    \centering
    \caption{Challenges for Blockchain-Based V2G Networks for Energy Trading}
    \label{table:energy-trading-BC-challenges}
    
    \scalebox{.7}{
    \begin{tabular}{|c|c|c|}
    \hline
        \thead{Challenge} &
        \thead{Description} & \thead{Blockchain-Based Solutions} \\
    \hhline{|=|=|=|}
    Identity privacy & \makecell[l]{Preventing leakage of EV identity\\information during communication\\ between V2G entities} & ~\cite{aggarwal1,baza2021privacy,9632819, MagazineAbouyoussef, 9500789} \\
    \hline
    Location privacy & \makecell[l]{Prevent leakage of EV location\\information during communication\\ between V2G entities} & \cite{corcoran1,sang1, li8, rasheed1, guo1, ying1, long1,yu1, singh3, benarous1} \\
    \hline Scalability & \makecell[l]{Supporting dynamic nature of EVs\\on network including increasing\\ number of transactions on network}  & \cite{hassija1,baza2021privacy}\\
    \hline
    \makecell[c]{Competitive bidding for \\energy prices} & \makecell[l]{Supporting competitive bidding for \\energy trading prices  over network} & \cite{pustisek1,knirsch1,kang1,su1,huang1} \\
    \hline
    \end{tabular}
    }
\end{table}

To solve the problem of identity privacy, the authors in \cite{aggarwal1} proposed a Blockchain-based efficient authentication scheme for V2G networks designed to preserve the identity of EVs.    Merkle Root Hash (MRH) is used to verify and validate transactions before adding new blocks to the Blockchains. The scheme also provides for a reward mechanism for EVs for participating in the regulatory and managing process. A major drawback with this scheme is that it is not fully decentralized as it requires utility company to act as the third party to maintain the Blockchain and validate transactions. In another works \cite{baza2021privacy,baza2021Blockchain}, the authors have proposed using pseudonyms to  preserve privacy, however because EVs are anonymous, a dishonest EV may abuse such anonymity by pretending as multiple non-exiting EVs to submit multiple reservations/offers without committing to them. To address this problem and  a common prefix linkable anonymous authentication scheme have been used so that if an EV submits multiple reservations/offers at the specific timeslot, the Blockchain nodes can identify and reject such submissions. In another works~\cite{9632819, MagazineAbouyoussef, 9500789}, Blockchain has been used  to enable Dynamic wireless charging of EVS to  enable the exchange of power between a mobile EV and the electricity grid via a set of charging pads (CPs) deployed along the road. To ensure EVs anonymity and  data unlinkability, group signature technique is used to enable EVs to send charging requests anonymously. However, there is no implementation to measure the scalabilty and performance issues of using Blockchain technology in the case of dynamic wireless charging of EVS.

To address location privacy, techniques employed in the research include obfuscation \cite{corcoran1,benarous2022obfuscation}, differential privacy \cite{sang1, li8, rasheed1,liu2022hut,li2022break}, K-anonymity \cite{guo1, ying1, long1,li2022quantifying, hu2022user} and pseudonym exchanges \cite{yu1, singh3, benarous1} and combination of all these techniques~\cite{de2022impact}. For example,  K-anonymity is a measure of anonymity of  entities within a dataset where methods such as suppression or generalization of data within that dataset are used to ensure the individuals or entities within that dataset cannot be reidentified \cite{samarati1}. Such reidentification can be either directly or indirectly by inference by comparing the individual's data to data within another dataset \cite{samarati1}. It is referred to as K-anonymity because there must be at least k other entities that share the same set of identifiable attributes for the entity to be considered k-anonymous. Methods for increasing K-anonymity include removing, hiding or encrypting identifiable attributes for the entity within the dataset. The authors in \cite{long1} suggest that K-anonymity best serves vehicle identity privacy in the context of Blockchain-based V2G networks. In this work, the authors propose a P2P EV energy trading scheme that makes use of consortium Blockchain and K-anonymity based on an undirected graph to protect the location privacy of EVs. EVs within the network can be visualized as nodes or vertices and the connections between the EVs as edges. In this work, not all EVs share connections with all other EVs.  When an EV makes an energy trading request, it collects the locations and identities of all other EVs on the network and sends the locations and identities of all these EVs in the request along with its own. By including all EVs within each request, an adversary would have a difficult time identifying which EV made the actual request.

Scalability of the V2G network is another challenge to integrating Blockchain in V2G network for energy trading. To address the scalability challenge, the authors in \cite{hassija1} proposed using a lightweight Blockchain-based protocol called directed acyclic graph  that uses an IOTA ledger with a tangle data structure to record transactions securely and in a scalable manner. The authors uses IOTA ledger  as is more suitable for recording and processing a large number of micro-transactions. Also, IOTA networks are less susceptible to distributed denial of service (DDoS) attacks since no single node retains the unique privilege to create or maintain the data structure. The work developed a tip selection algorithm that allows the addition of new transactions without the need of miners, thereby decreasing the amount of computation power needed. Here, a tip refers to a transaction that has yet to be approved. Game theory is used  to allow EVs negotiate with the grid for the best price and match EVs with the most optimal grid.

In energy trading one of the concerns is to select an EV with the best bid price. Authors in \cite{pustisek1}, has used smart contract to select the most appropriate bid price dynamically among all the charging station where Blockchain is used to maintain information about the EV route, car battery status and the preference of driver. After getting updated information from Blockchain, charging stations broadcast their desired price to the EV network from where EV selects the best one though the underlying payment system is still one of the limitations of their proposed scheme.  In \cite{knirsch1}, a hidden commitment is sent to the Blockchain by the EV for reserving a place in the charging station with the best price though management of this system is not good as the charging station does not aware about the reservation until the EV reaches the definite charging station. Also, authors do not mention about the payment system along with the privacy protection of the EVs as well as this proposed scheme is prone to Sybil attacks and denial of service (DoS) attack. Optimization for the electricity consumed and its price is also a concern issue in energy trading which is mentioned in \cite{kang1} where authors have proposed a decentralized electricity trading system based on consortium Blockchain by using iterative double auction optimization method or V2V charging.  In energy trading it is also needed to maximize the operator\textquotesingle s profit while to give the EV best service price. To make the operator\textquotesingle s benefit maximum, authors in \cite{su1}, has proposed a contract theory with integration of Byzantine fault tolerance consensus algorithm to reach the consensus in the permissioned energy Blockchain. Another scenario to maximize user\textquotesingle s benefit along with the minimizing the cost has been proposed in \cite{huang1}, where authors have utilized a double- objective optimization model for constructing this optimal charging scheduling algorithm that considers both V2V and CS2V.

Wu et al. \cite{wu1} discuss how to keep a P2P energy trading secure and reliable as new roles and players such as EVs are injected into energy systems. They also discuss how to improve the cooperation between regulated energy system players such as utility companies and non-traditional players such as EVs. The authors suggest that blockchain and multi-scale flexibility services can serve as an key enablers for solving issues of security, reliability and trust and better coordination between traditional and non-traditional energy-trading roles. Smart contracts can be employed as self-regulating and self-executing mechanisms for matching energy producers with consumers for energy trading. Smart contracts can also be implemented to automate flexibility processing, including the clarification of energy trading roles (providers and suppliers) and responsibilities. One drawback to their proposal is that it is not completely decentralized as they suggest professional players and decision makers should play the role of regulating the energy system.

As mentioned, EVs can sell energy in exchange for rewards or payment. However, a malicious server may attempt to develop behavior profile for an EV, which poses a privacy issue. To solve this issue, one could apply a private data sharing scheme similar to the one proposed by \cite{li11}. Although this proposal focuses on securing private data on IoT devices in general, a similar scheme could be applied to EV private data concerning energy trading. The authors proposed a blockchain-based privacy preserving and rewarding private data sharing scheme for IoT. The authors make use of deniable ring signature and Monero to implement behavior profile building prevention with non-frameability. The proposal also uses licensing technology executed by smart contracts to ensure flexible access control for multi-sharing.

\subsection{Crowdsourcing-based applications}

Crowdsourcing is a novel model in which customers as well as organizations obtain goods or services. Examples of Crowdsourcing includes ride sharing and smart parking systems.   During ride sharing one of the major concerns is to protect the privacy for both riders and drivers and riders as for accessing this service they need to share their personal information such as location. To solve this problem, the work in \cite{li3} has proposed a Blockchain-based zero-knowledge proof protocol by which a rider can his service from a verified driver that will be done by the peer nodes in the Blockchain. Tough this problem does not address the problem associated with the attacker disguised as the rider and also due to presence of a central authority for issuing the keys, this model is not fully decentralized.
 
Computational overhead is a common drawback in BIoV during ridesharing. To solve this issue, in \cite{wang1}, authors have proposed consortium Blockchain which uses attribute-based proxy re-encryption algorithm for guaranteeing the security and privacy of the shared data. The DPoS (Delegated Proof of Stake) is used here for storing the block and verifying it for traceability if needed. On the other hand, fog computing is used in \cite{li5}, to match drivers with the passengers locally which can lessen the computational overhead. Authors here used a private proximity test with location information for generating a unique secret key which is used between passenger and driver. A private Blockchain has also been introduced for keeping the record of ridesharing. Here, hash value is stored in Blockchain whereas encrypted data is stored in cloud.

To address security and privacy issues in ride sharing, In~\cite{baza2019b}, Baza et. al have proposed a privacy-preserving ride sharing scheme using the permissionsless Blockchain. The main idea of the proposed scheme is to that riders hide their exact pick-up and drop-off location information by cloaking their location information. The authors also addressed the problem of malicious behaviour of drivers and riders using the open and permissionsless Blockchain. The main idea is that dishonest drivers and riders are published by  paying an initial deposit before the beginning of the trip. Then, using zero-knowledge proof, drivers have to prove to the Blockchain their good willing by submitting the pick-up location. The Blockchain can verify that the driver lies in a set of predefined pick-up locations without knowing the exact pick-up location to preserve privacy. The proposed techniques use the IoV infrastructure so RSUs verify the driver's pick up location before submitting it to the Blockchain. In another work by~\cite{badr}, and instead of the cloaking technique, the authors proposed  to represent the ride sharing area into overlapping grids so  nearby riders/drivers can share rides. Then, for privacy-preservation, drivers/riders should encrypt their requests/offers using a lightweight cryptosystem, and then the Blockchain matches the encrypted offers and requests to find the best nearby driver/rider. 

Another sourcing models including smart parking system to enable individuals to share available parking slots to drivers.  Because the number of vehicles in the IoV is increasing exponentially which leads to a fast-growing problem of finding vacant parking slots. Latest parking system have several privacy and security vulnerabilities such as they are centralized which cause single point of failure. They also do not consider the privacy issue of the drivers \cite{salman1, lubin1, mazlan1, odonoghue1}. It is also needed to share the location information for this smart parking, and it is a big concern to provide location privacy during traffic data sharing or ridesharing
\cite{prakash1, hussein1, mosakheil1, radanovic1}. Blockchain is key to solve these problems regarding security and privacy. Authors in \cite{aljaroodi1,9045674}, proposed a Blockchain based smart parking system which can preserve the driver’s location by using the private information retrieval (PIR) technique. A short randomize signature is used to allow the drivers to put parking reservation request anonymously without exposing their real identities. Payment system here utilizes time locked criteria anonymously by exploiting smart contract and Blockchain between drivers and parking station.

\subsection {Vehicle Platooning}

\begin{figure}[!t]
    \centering
    \includegraphics[width=1\linewidth]{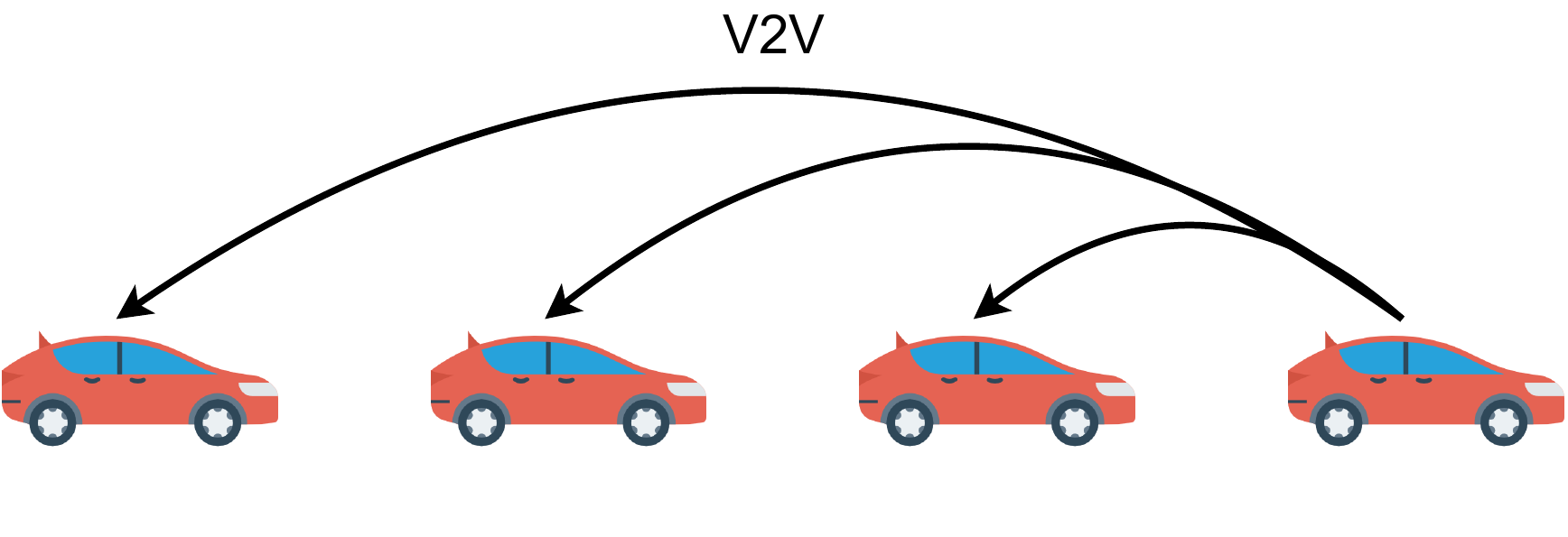}
    \caption{Vehicle Platooning. Lead vehicle transmits its velocity and acceleration to follower vehicles so that they can match the leader vehicle and maintain a safe distance between themselves}
    \label{fig:vehicle-platooning}
\end{figure}

The ever-increasing number of vehicles on the road has led to increased pollution \cite{jia1} and traffic congestion \cite{kaiwartya1, qiu1}. One potential solution to these problems is to utilize vehicle platooning. In a vehicle platoon system, a leader vehicle transmits its velocity and acceleration to other vehicles that follow it in the platoon. Based on the information transmitted from the platoon leader, the follower vehicles will, in turn, adjust their velocity and maintain a safe distance between themselves \cite{dey1, bashiri1, nardini1, ou1}. Figure \ref{fig:vehicle-platooning} illustrates a typical vehicle platoon system. In essence, the vehicles act similar to a train where the leader can be likened to a locomotive and the followers to the cars it pulls. Advantages of vehicle platooning include improved traffic capacity, traffic congestion reduction, reduced energy consumption and improved driving experience for platoon followers \cite{turri1, guo3, chen1}. For the platoon to successfully function, reliable and low-latency V2V communications between the vehicles are required. However, V2V communications can suffer from security and privacy issues. The security and privacy features offered by Blockchain could be used to solve such issues. Issues regarding implementing Blockchain in a vehicle platooning system include encouraging vehicles to serve as platoon leader, ensuring platoon leaders are paid for leading, and the handling the dynamic nature of vehicles entering and leaving platoons during their journey.

Proposals for integrating Blockchain into vehicle platooning systems exist as early as 2018. Wagner and McMillin \cite{wagner1} were the first to discuss Blockchain's potential role in vehicle platooning. The article's main focus is on securing IoVss using Blockchain but they discuss vehicle platooning as a use case and they provide a novel proposal for validating integrity of vehicles participating in the platoon. Ying et al. \cite{ying2} proposed a Blockchain-based platoon management system for autonomous vehicles. To increase the overall safety of the system, their proposal limits communication outside of the platoon to the platoon leader. The proposal also makes use of smart contracts for paying the platoon leader. The authors in \cite{chen1} proposed a vehicle platooning system that employs both smart contract and Blockchain for secure payment of platoon followers to the platoon leader. The authors' proposed system is designed to handle the dynamic nature of vehicles entering and leaving platoons throughout their journey. Their proposal utilizes path information to match the vehicles and elects platoon leaders based on their recorded reputation. The authors in \cite{li10} proposed a truck platooning system with a location-aware and privacy-preserving verification protocol based on zero-knowledge proof and permissioned Blockchain. Their proposed system provides lower latency and communication overhead by performing the verification process within a spatially-local area defined by the platoon.

\subsection{Traffic Congestion Reduction}

IoV implementations of IoVs can serve to reduce traffic congestion on roads and highways. Information regarding traffic congestion can be shared between vehicles~\cite{vishwakarma2022smartcoin,diallo2022blockchain}. Vehicles who have yet to approach the congested area can use this information to alter their routes and avoid the congestion. This, in turn, will help reduce the congested area. However, the trading of road congestion information raises security and privacy concerns especially in terms of revealing the location information of vehicles. Features of Blockchain could be utilized to solve these security and privacy concerns~\cite{baker2022blockchain,dungan2022blockchain}.

Li et al. \cite{li9} proposed a trust management protocol called \textit{CreditCoin} which makes use of an incentive mechanism to encourage vehicles to share traffic information. Under authors' proposal, RSUs and official public vehicles make use of authors' proposed BFT-based consensus mechanism, which repeats a voting process until an agreement is reached. Advantages of authors' proposal include low computation time and the ability to defend against Sybil attacks. Drawbacks to authors' proposal include limited scalability and privacy leakage concerns. Although BFT-based consensus provides overall good performance, it only provides limited scalability \cite{vukoli1}. Privacy leakage concerns exist in authors' proposal because the system combines credits with transactions and transactions are accessible by everyone. Liu et al. \cite{liu1} proposed a Blockchain-based trust management model with a conditional privacy-preserving announcement scheme for IoVs. The system uses an anonymous aggregate announcement protocol to allow vehicles to send messages such as traffic information anonymously and thereby guarantee the privacy of the vehicles in an untrusted IoVs. Similarly, Yang et al. \cite{yang1} proposed a decentralized trust management system for IoVs using Blockchain where vehicles can validate messages containing information such as traffic information from other vehicles using a Bayesian Inference Model. Vehicles rate each vehicle using this model and upload their ratings to RSUs, which are responsible for calculating the trust value for each vehicle and packaging such values into a block in Blockchain. Their system makes use of a combination of PoS and PoW consensus algorithms. One drawbacks to this work includes the potential for privacy leakage related to the fact that authors' proposal uses vehicle identity number and a malicious entity could use this information to discover the identity of the vehicle. The authors suggest a future work could make use of public key encryption to solve this problem.

Though our current traffic can be turned into decentralized by adding Blockchain, then this system will face a new challenge associated with data integrity and data privacy during the incorporation of connected vehicle data into real traffic system because of the transparency advantage in Blockchain. In \cite{li1}, authors propose a location aware traffic management with addition of non-interactive zero knowledge range proof protocol (ZKRP) and permissioned modular Blockchain which is based on Hyperledger fabric platform and Hyperledger Ursa cryptographic library. This data management which has also a gateway that authenticates each incoming vehicle. Instead of providing the incoming vehicle’s identity, it will send a ZKRP encrypted message to the gateway for validation where the gateway acts as the verifier and the incoming vehicle acts as the prover. This model can stand against spoofing attack, eavesdropping attack and tampering but it requires high running time required as Hyperledger Ursa does not compute in parallel and this model doesn't provide any scenario if the gateway encounters more than one request at a time in multiple region scenario.

Kumar et al. \cite{kumar1}, employ Blockchain along with deep learning to address security and privacy vulnerabilities in a cooperative intelligent transportation system (C-ITS). The proposed framework addresses  security issues using Blockchain and deep learning modules. The Blockchain module is used to securely transmit data between AVs, RSUs and traffic command centers. The module uses a smart contract-based enhanced PoW (ePoW), a consensus algorithm first proposed by Keshk et al. \cite{keshk1}, which is more energy-efficient alternative to PoW and PoS and is designed to authenticate data records and prevent data poisoning. ePoW estimates its proof based on the number of records in the dataset. Lastly, the proposal by Kumar et al.'s \cite{kumar1} has a deep learning module that uses a Long-Short Term Memory-AutoEncoder (LSTM-AE) to encode C-ITS data to prevent inference attacks. The authors propose an Attention-based RNN (A-RNN) to analyze the encoded data and detect intrusive events on the C-ITS.

\subsection{Collision and Accident Avoidance}
\label{sec:collision-avoidance}

AVs use non-linear controllers in order to maneuver through the dynamic and ever-changing nature of vehicles on the road \cite{hamid1}. Vehicle-to-everything (V2X) communication can be harnessed to share information gleaned from on-board sensors with other vehicles and alert them of objects or persons on the road and other hazardous road and environment conditions so that vehicles can take appropriate actions to avoid collisions with objects or persons or accidents \cite{abboud1}. The increased connectivity and reliability provided by IoV can further improve this feature. V2X refers to communications between vehicle and things within its environment such as traffic lights, pedestrians and other vehicles. As such, V2X is also an all-encompassing term that includes other communications such as vehicle-to-infrastructure (V2I), vehicle-to-RSU (V2R), vehicle-to-onboard-sensors (V2S) and vehicle-to-vehicle (V2V) communications. The increased connectivity and reliability provided by IoV can provide faster and more efficient delivery of V2I communications regarding the road environment to vehicles which, in turn, can serve to enhance vehicle localization. Localization refers to the method by which an IoV determines the accurate location of a vehicle. However, with such increased localization also arise concerns of security and privacy of vehicle location. One should also consider the security and privacy of other entities participating over IoV. The security and privacy features of Blockchain could be harnessed to solve such issues.

Robin Westerlund \cite{westerlund1} proposed an Ethereum Blockchain-based system where AVs use smart contracts to reserve and exchange road spaces for travel. According to author, this system could be used to ensure AVs do not try to occupy the same spaces and maintain a safe distance from one another so as to prevent accidents and collisions. Through a simulation, the author demonstrated that the proposed system provides a high level of security and reliability.

Another approach is to apply the concept of swarm robotics to the cooperation of AVs within the context of IoVss and IoV. Under swarm robotics, robots communicate and share information with the goal of coordinating actions for a common goal. Based on this concept, Singh et al. \cite{singh4} proposed a Blockchain-based framework using PoA as an efficient consensus mechanism for swarm robotics. A similar swarm technique could be explored in the context of AVs within a IoVs or IoV. Swarm coordination of AVs secured by Blockchain could be used to coordinate the movement of vehicles on the road to prevent collisions and, in particular, the coordination between AVs crossing through an intersection.

IoV can also be harnessed to reduce collisions resulting from improper lane changing decisions and those caused by failure to identify vehicles, particularly smaller vehicles such as motorcycles, approaching a vehicle from its blind spot. The authors in \cite{fu2} proposed a Blockchain-based collective learning (BCL) framework for the secure and efficient lane changing of connected and autonomous vehicles (CAVs). BCL adopts federated learning (FL). Using FL in lieu of a central server solves the problem of reliance on a central server and single-point-of-failure (SPOF) attacks and allows vehicles to take advantage of the collective intelligence of all vehicles on the network. Using consortium Blockchain to store the collective learning regarding lane-changing behavior further secures the learning process and helps prevent false information sharing.

Buzachis et al. \cite{8969653} proposed an autonomous intersection management system that uses Hyperledger Fabric permissioned Blockchain and smart contracts to securely manage AVs as they cross through intersections. AVs use V2I communication to relay information such as vehicle identity, speed and position to an intersection management agent. To ensure only valid AVs make crossing requests with the intersection management agent, AVs are required to send their requests using digitally signed messages using public key encryption. The intersection management agent deploys a smart contract with the AV how to safely cross the intersection and records each crossing in Blockchain. In the case of multiple requests from different AVs to cross an intersection, the system uses a first-come first-serve (FCFS) algorithm to determine the order of intersection crossings. One drawback for authors' proposal is that it is not completely decentralized as intersection management agents are assumed to be under the auspices of public traffic authorities.

In a later work, Buzachis et al. \cite{buzachis2} propose a similar intersection management system based on the agreement between AVs through V2V communication rather than relying on an intersection management agent. Similar to \cite{8969653}, the system makes use of Hyperledger Fabric permissioned Blockchain and the first-come first-serve algorithm. However, in this proposal, AVs host the Blockchain ledgers and conduct smart contracts. Based on simulations, however, the authors concluded that the relay of information between AVs was too slow to provide reliable real-time intersection passing decisions. Future research can be done to decrease communication latency and make the system more scalable with a larger amount of AVs requesting passage.

\subsection{Infotainment and Content Cashing}

Implementing a IoVs using IoV improves better Internet connectivity for vehicles and this increased connectivity can serve to improve the delivery of in-vehicle infotainment to vehicles such as music and video content for passengers. Such content will enhance the experience and enjoyment of drivers and passengers. However, this connectivity raises concerns regarding the security and privacy of the content delivered to vehicles. The increased security and privacy provided by Blockchain could be used to solve these concerns. Kim and Ryu \cite{kim2} proposed a system called \textit{autoCoin} designed to encourage smart vehicles to exchange infotainment securely. Their system makes use of a scalable Blockchain architecture and smart contracts to exchange infotainment contents in a peer-to-peer fashion rather than rely on a third-party to deliver infotainment content. In addition, \textit{autoCoin} makes use of a variant of PoS called Proof-of-Stake Deposit (PoSD) as a consensus mechanism and includes a reward mechanism to encourage vehicles to share content.

\begin{figure}
    \centering
    \includegraphics[scale = 0.45]{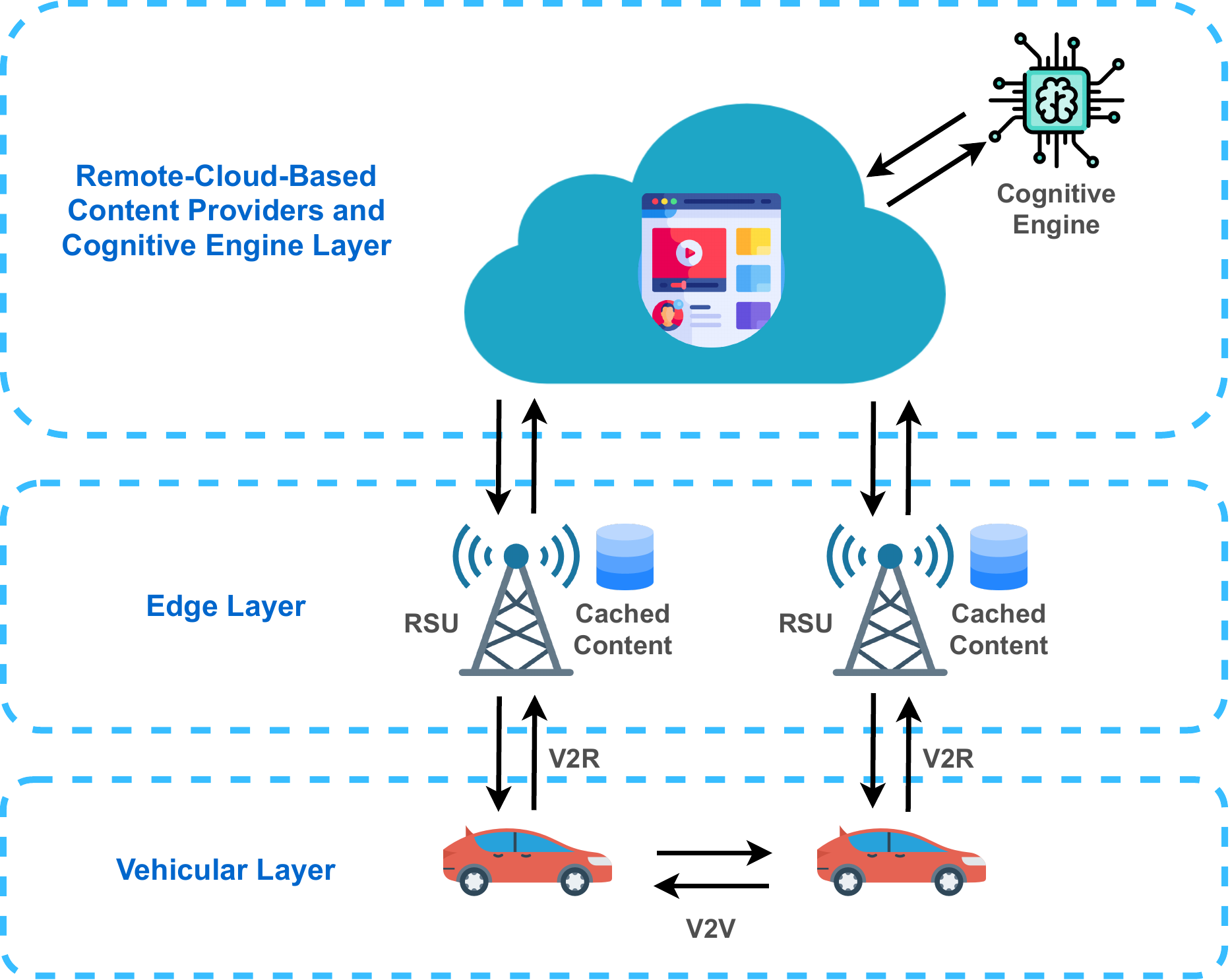}
    \caption[Short caption]{Traditional CIoV Infrastructure}
    \label{fig:traditional ciov infrastructure}
\end{figure}

Some propositions for IoV utilize a cognitive engine to perceive content requirements of users and match them with content providers and thereby improve the content caching hit rate. Figure \ref{fig:traditional ciov infrastructure} illustrates a traditional cognitive IoV (CIoV) infrastructure. A CIoV infrastructure consists of three layers: (1) vehicular layer, (2) edge layer, and (3) remote-cloud-based content providers and cognitive engine layer. In the vehicular layer, vehicles collect data using onboard sensors and send them to the cloud for data analysis by the cognitive engine. Vehicles in this layer also communicate with each other using V2V communication. In the edge layer, vehicles communicate with RSUs using vehicle-to-RSU (V2R) communication. In the remote-cloud-based content provides and content cognitive engine layer, a cognitive engine percieves the needs of vehicles using deep learning and ML.

\subsubsection{CIoV Challenges and Solutions}
There are several issues with CIoV~\cite{li2022blockchain,thai2022design,wadhwa2022heterofl}. Table \ref{table:traditional ciov infrastructure challenges} summarizes the various challenges to implementing an effective CIoV infrastructure. First, the process used by a CIoV introduces privacy and security issues. Contents requesters may submit data such as points of interest or other private data that could be leaked. Also requesters may receive data from untrusted parties. Other problems with traditional CIoV include limited transaction time for obtaining contents, the cognitive engine cannot accurately determine true content requirements, the content stored on nearby RSUs may not meet the needs of a requester, and the constant broadcast of content

\begin{table}[h]
    \centering
    \caption{Challenges for Traditional CIoV Infrastructures}
    \label{table:traditional ciov infrastructure challenges}
    \begin{tabular}{|l|l|}
    \hline
        \thead{Challenge} &
        \thead{Description} \\
    \hhline{|=|=|}
        \makecell[l]{Privacy} & \makecell[l]{Content requesters may\\submit private data that\\could be leaked such as\\points of interest.}  \\
    \hline
        \makecell[l]{Security} & \makecell[l]{Content requesters may\\receive data from\\untrusted parties.}\\
    \hline
        \makecell[l]{Limited Transaction Time} & \makecell[l]{Limited transaction time\\for obtaining contents} \\
    \hline
        \makecell[l]{Failure to accurately determine\\content requirements of requestors} & \makecell[l]{Cognitive engine may not\\accurately determine the\\true content requirements\\of content requesters} \\
    \hline
        \makecell[l]{Failure to store proper contents\\on nearbye RSU} & \makecell[l]{CIoV framework may not\\store content on nearby\\RSUs that meet the needs\\of the content requesters} \\
    \hline
        \makecell[l]{Constant broadcast of contents} & \makecell[l]{Requiring content be\\constantly broadcasted\\imposes high storage\\and communication costs} \\
    \hline
    \end{tabular}
\end{table}

\begin{figure}
    \centering
    \includegraphics[scale = 0.55]{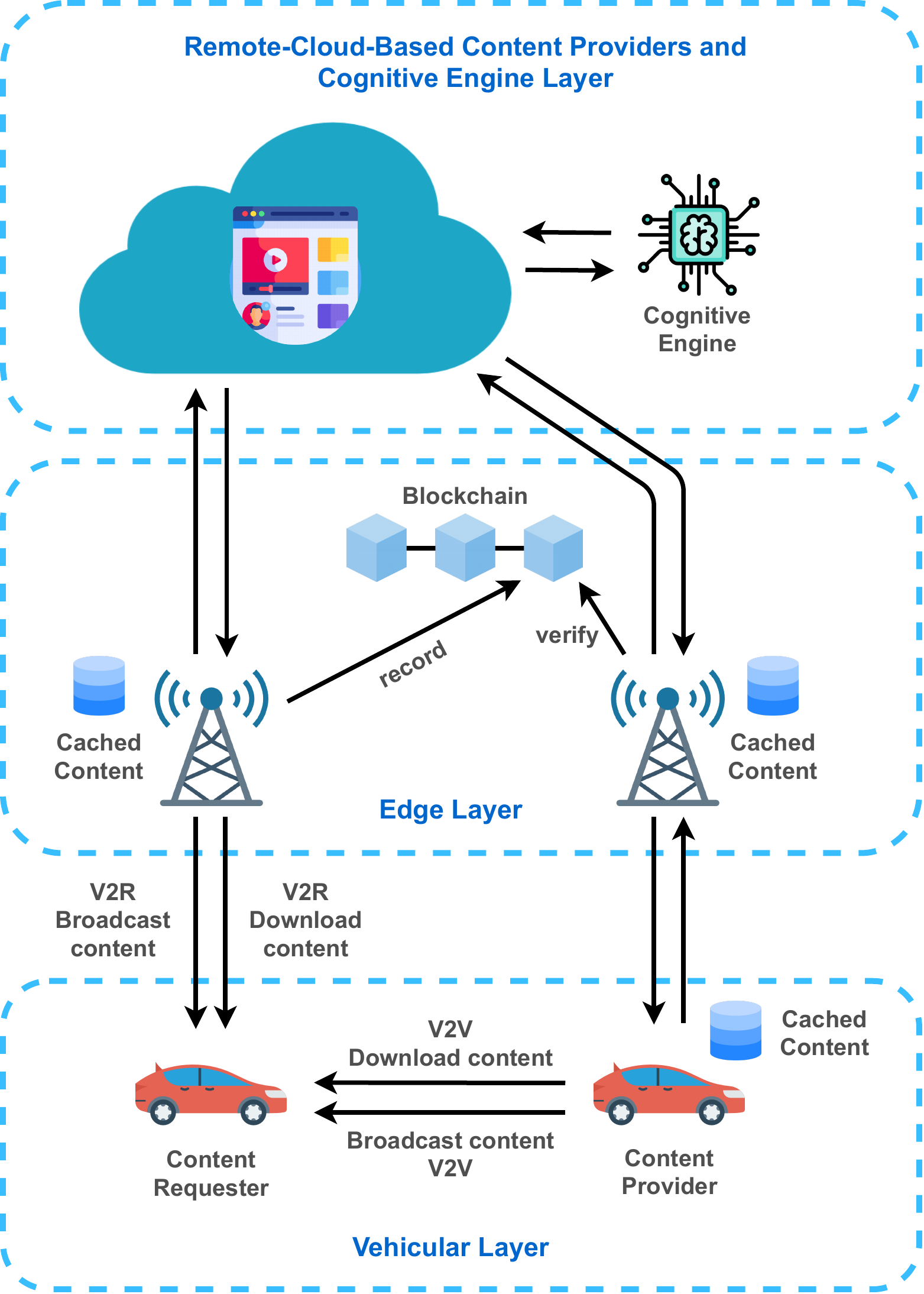}
    \caption[Short caption]{Blockchain-based Privacy-Aware Content Caching Architecture. There are four phases: (1) cognitive engine data analysis in which the cognitive engine sends recommended content sets to vehicles.  The cognitive engine uses  data analysis and AI techniques to update the content caching and improve the content caching hit ratio. (2) Vehicle registration phase in which each vehicle chooses a role as either content requester or content provider. A provider can also change its role to requester when it wants to request content  (3) content broadcasting and acquisition in which a provider broadcasts content and the required download time for the content. and (4) content transaction in which each completed content sharing transaction is recorded by the RSU and stored in the Blockchain.  
}
    \label{fig:BC-based privacy-aware content caching architecture}
\end{figure}

In response to these problems, the authors in \cite{qian1} proposed a Blockchain-based privacy-aware content caching architecture in CIoV. Figure \ref{fig:BC-based privacy-aware content caching architecture} shows the architecture of the proposed scheme. The work aims to improve the cache hit ratio by designing the cognitive engine to perceive the content requirements of requesters using ML or deep learning and sending such perceived content requirements to RSUs ahead of time. Vehicles receive content by downloading it when RSUs and vehicles broadcast it. Since content requester vehicles just receive the broadcast, rather than sending an individual request that contains the requested content, the requested content is private.  A cognitive engine is responsible for detecting when content requesters connect to providers by V2R or V2V communication and record the transaction. Providers in authors' scheme need to collaborate together to provide content because transient vehicle requesters may not always be in range of any one provider. When a provider provides content, it receives a reward and after the transaction is complete, the cognitive engine will record it in Blockchain.

AI and ML technologies can also be employed to enhance and optimize the delivery of infotainment to vehicles for Blockchain-enabled IoVs. For example cognitive engines can be employed to intelligently perceive content requirements of users, match them with content providers and cache them at RSUs. Qian et al. \cite{qian1} proposed a Blockchain-based privacy-aware content caching architecture that records content sharing transactions in Blockchain. They also designed a cognitive engine for their proposal that utilizes ML to perceive content requirements and store them in RSUs ahead of time. Similarly, Dai et al. \cite{dai1} proposed a Blockchain-empowered distributed caching framework where content is cached in participating vehicles. Base stations serve to pair caching requesters to caching providers and use AI to predict V2V transmission ranges and connection duration between content requesters and providers. 

\subsection{Summary Remarks}

In this section, several BIoV applications have been discussed in details.  Blockchain is a promising solution to solving many of the problems  security of V2V communications in energy trading, crowdsourcing, traffic congestion and collision avoidance,  vehicle platooning and infotainment applications.  Trust management systems that utilize Blockchain are a promising solution to solving issues related to the security and privacy of vehicles when sharing traffic-congestion information. The reputation of vehicles sending accurate traffic information can be stored in Blockchain and used to limit the possibility of vehicles sending false traffic information and such systems can be supplemented with incentive mechanisms to encourage the sharing of traffic information. Future work needs to be done to solve the issues of privacy leakage and selecting the most efficient consensus mechanism that is both capable of handling the dynamic nature of the road and ensuring scalability of the IoVs.

 Using Blockchain and smart contracts can be utilized to aid vehicles to securely coordinate their positions and share information about road hazards. ML and AI technologies can be harnessed to supplement the system, giving vehicles access to collective learning of all vehicles to help them make more intelligent decisions with regard to lane changing decisions. One of the main issues is scalability and performance issues of using Blockchain technology in the case of dynamic wireless charging of EVS. As we discussed, dynamic wireless of EVs enables the exchange of power between a mobile EV and the electricity grid via a set of charging pads (CPs) deployed along the road.  Using Blockchain can be used to provide secure dynamic wireless charging of EVs, however because of scalability and performance issues of current Blockchains, there is more research is required to provide scalable and fast charging using Blockchain.

\section{Federated Learning for BIoV} \label{sec:BC FL for IoVs}

Federated Learning (FL) is a distributed machine learning (DML) approach that has garnered recent increased interest due to ability to learn from data without sharing the data. However, some challenges for FL include defending against poisoning and member inference attacks and centralization and loyalty issues. In response to these challenges, there have been a number of proposals to enhance traditional FL with Blockchain.

In this section, we first distributed learning followed by an  an overview of FL. Then, we discuss challenges with existing FL approaches, how blockchains can resolve these challenges, and existing blockchain-based FL for IoV applications. Finally, we summarize of the applications of Blockchain in FL for vehicular networks and some of the existing challenges in these approaches.

\subsection{Overview of Distributed Learning \& Federated Learning}
\label{DML overview}

In traditional ML, a single computer or central server trains using locally stored data \cite{verbraeken1, qiu2, zheng2, coulouris1, sherif1, sherif2, sherif3}. This approach lacks scalability, consumes more time and is not feasible for big and distributed data applications that require real-time decisions \cite{qiu2, peteiro1}. Scalability is the ability to manage increases in computation load or in the number of users \cite{coulouris1}. If the ML training data gets too large, it may exceed the memory and computational resources of the central server \cite{peteiro1}. A good training model requires a great deal of training data. Failure to obtain a sufficient amount of training data could result in an inaccurate global model.

In response to these limitations, DML was developed, where learning models can be trained on subsets of data distributed across several nodes \cite{peteiro1, zheng2}. The primary motivation behind DML arises from the need to share resources, such as the computing abilities of multiple computers \cite{coulouris1} and the distributed nature on current networking applications. The advantages for training across multiple nodes include greater scalability and efficiency. Parallel execution of learning tasks across multiple nodes leads to reduced training time. DML provides a higher degree of availability, which is the measure of the proportion of time a system is available for use\cite{coulouris1}. The distributed nature of the nodes solves problems associated with running out of computing, storage and memory resources on a central server.

Despite the many advantages DML provides, it still suffers from issues in heterogeneity, security, and failure handling and it needs improvements to its scalability \cite{coulouris1}. DML suffers from heterogeneity: A DML may find it difficult to train across multiple diverse nodes each with differing memory, storage and computation abilities and resources \cite{coulouris1}. Security and privacy issues arise when vehicles transmit raw data collected from their sensors \cite{kang2}. For example, raw data may contain private information such as the location or destination of a vehicle. DML also suffers from failure handling, requiring measures to manage situations when one node fails \cite{coulouris1}.

\begin{figure}[!t]
    \centering
    \includegraphics[scale = 0.6]{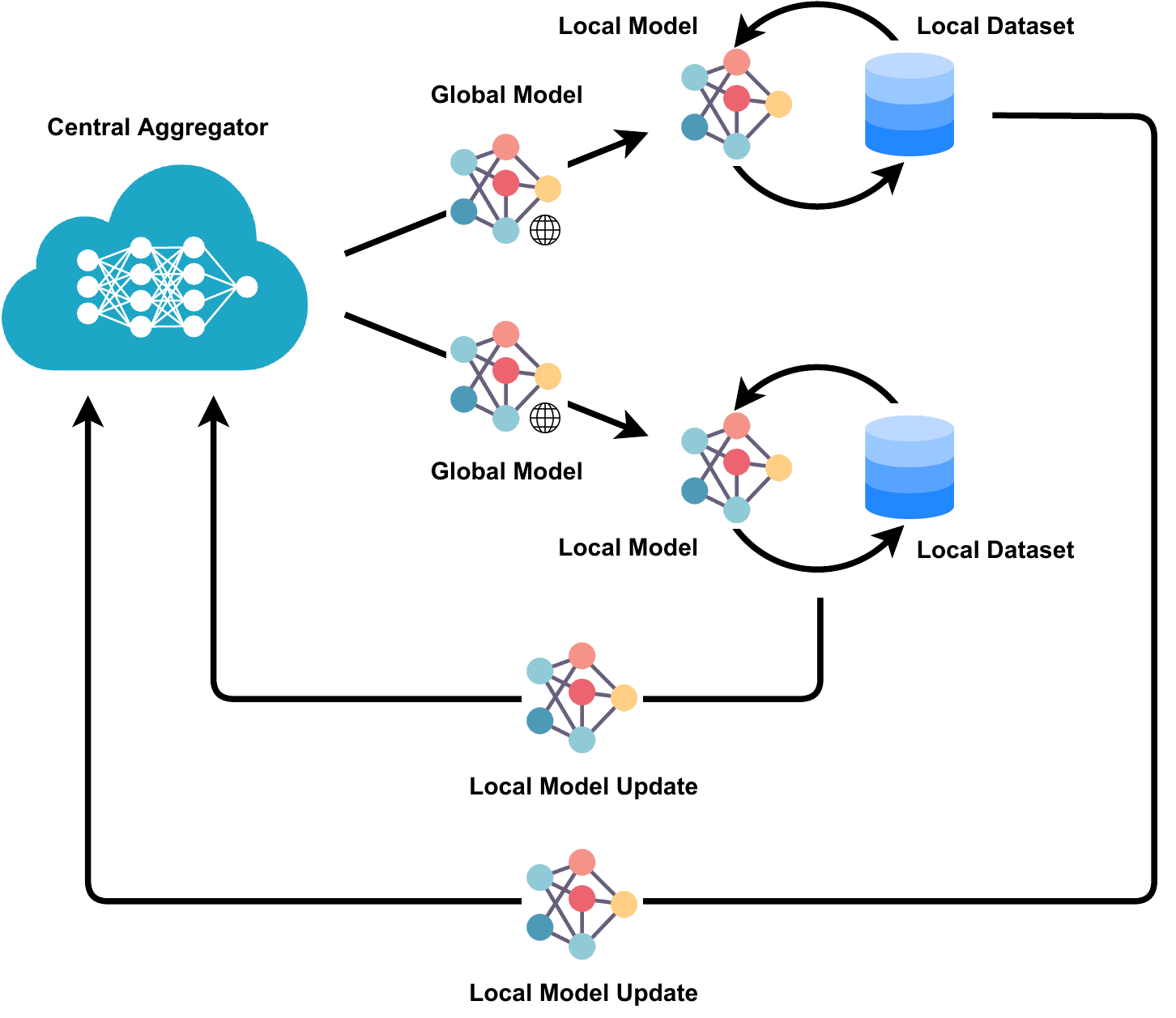}
    \caption[Short caption]{Traditional Federated Learning Framework}
    \label{fig:traditional FL framework}
\end{figure}

In response to above concerns, researchers at Google Inc. \cite{mcmahan1} proposed and developed FL as a new approach to DML. FL involves training distributed local models on local devices or vehicles (referred to as clients or participants) in a federated fashion \cite{mcmahan1}. FL consists of three primary phases: initialization, aggregation, and update. Based on general descriptions in the literature, we created an illustration of traditional FL framework shown in Figure \ref{fig:traditional FL framework}. During initialization, FL participants download a global model from a central aggregator. Each participant uses this global model as the basis for training their own new local model using locally collected data. When finished training their local model, the participants upload a local model update to the central aggregator which is responsible for aggregating the local models of all participants to form a new global model. Only the local model update are transmitted. The training data itself remains with the participant. After aggregation, the participants download and begin training on the new model. This process repeats for multiple iterations until the global model reaches a preset accuracy or optimal convergence. 

The primary difference between FL and DML is that DML focuses on harnessing parallel computation while FL concerns training models locally on each client \cite{parimala1}. DML assumes that datasets are distributed in identical sizes across all nodes. In FL, however, the size of the datasets for each node can be heterogeneous or of differing sizes \cite{sattler1}.  FL offers greater privacy than the traditional ML on a central server. Participants train locally using their own collected data. FL provides greater privacy by only requiring participants share local model updates with the central server rather than raw data \cite{kang2}. Due to the advanced privacy offered by the FL approach, many researchers have adopted FL for applications such as healthcare data \cite{xu1, huang2}, medicine \cite{sheller1}, smart city sensing \cite{jiang1}, smartphone next-word prediction \cite{hard1}, and improving Google keyboard suggestions \cite{yang2}.
 
\subsection{Challenges for Traditional Federated Learning} \label{FL challenges}

Despite the privacy enhancements FL offers, FL suffers from challenges such as centralization, loyalty, vulnerability to single-point-of-failure (SPOF), poisoning and membership inference attacks, and communication efficiency \cite{pokhrel1, kang3, chai1, qi1, liu2, mcmahan2, bonawitz1, lim1, zhu1, wei1, tian1}. Table \ref{table:traditional FL challenges} provides a summary of these challenges and discussed in details as follows. 

\begin{itemize}

\item \textbf{Centralization Issues}. This problem exists because FL relies on a central server to aggregate local model, i.e., the aggregator \cite{pokhrel1}. If the aggregator malfunctions or is compromised e.g., by a SPOF attack, the entire FL system could fail. A \textbf{\textit{single-point-of-failure attack}} is an attack designed to take down any network element, such as a central server in FL, that would cause the entire network to fail.

\item  \textbf{Lack of loyalty mechanism.} Traditional FL also lacks a loyalty mechanism for encouraging participants to share their local learning models. Some works, such as \cite{kang3}, have suggested using a contract theory approach to design an incentive mechanism to encourage participants to participate in FL. Also, the authors of \cite{chai1} utilize a trading market process to stimulate the sharing of learning models.

\item  \textbf{Poisoning Attack.} This attack occurs when a participant submits false or low-quality model updates that "poison" the FL model, which can result in a less accurate global model or cause the global model to outright fail \cite{qi1}. 

\item  \textbf{Membership inference attack.} This attack occurs when an adversary utilizes reverse engineering techniques to gain access to a participant's local model updates from the central server and uses this information to steal the local dataset of the participant \cite{liu2, qi1}.

\item  \textbf{Communication efficiency}. Communication remains a critical bottleneck in FL networks \cite{bonawitz1}. The vast number of devices or vehicles on an FL network combined with their limited resources such as bandwidth, energy, and power can slow down communication across the FL network. For example, vehicles with greater computation resources would still have to wait on those with less resources.
To address this challenge, many researchers have focused on reducing the number of communication rounds \cite{mcmahan1, li4} or decreasing the transmission payload size \cite{tian1, zheng1, reisizadeh1, du1}. According to \cite{wei1}, these approaches assume a noise-free communication scenario and focus primarily on ML design that trades off computation and communication. More recent works focus on communication system design in lieu of ML design \cite{lim1, zhu1, wei1}. For example, Wei and Shen \cite{wei1} investigated how to improve communication performance when communication occurs over noisy channels or those that contain communication errors.

\end{itemize}

\begin{table}[!t]
    \centering
    \caption{Challenges for Traditional FL}
    \label{table:traditional FL challenges}
    \scalebox{0.9}{
    \begin{tabular}{|l|l|}
    \hline
        \thead{Challenge} &
        \thead{Description} \\
    \hhline{|=|=|}
        \makecell[l]{SPOF attack\\(centralization)} & \makecell[l]{FL relies too heavily on central server.\\Any attack on central server may\\cause entire network to fail.} \\
    \hline
        Lack of loyalty mechanism &
        \makecell[l]{FL lacks loyalty mechanism to encourage\\participants to share their local models.}\\
     
     \hline
        Poisoning attacks & \makecell[l]{Participant submits false or low-quality\\ model updates that "poison" FL model.\\Can result in a less accurate global\\ model or cause the global model to fail} \\
    \hline
        \makecell[l]{Membership\\inference attack} & \makecell[l]{
        Adversary gains  access to a  participant’s\\local model updates from central server\\and uses this information to steal\\the local dataset of the participant}\\
    \hline
        \makecell[l]{Communication\\efficiency} & \makecell[l]{Large number of devices or vehicles on\\FL network combined with their limited\\resources such as bandwidth, energy, and\\power can slow down communication\\across FL network} \\
    \hline
    \end{tabular}
    }
    \label{tab:my_label}
\end{table}

\subsection{Federated Learning in IoV Enhanced by Blockchain} \label{BC-enabled FL for IoVss}

Blockchain's distributed and decentralized features have the potential to solve the challenges of traditional FL. For example, replacing traditional FL's central server with Blockchain solves the centralization problem and protects against SPOF attacks. Verification Procedures can be implemented into Blockchain aggregation to help protect against data poisoning attacks. To solve problems with loyalty, procedures can be implemented into Blockchain such as providing rewards for vehicles or miners for supplying good model updates.

\begin{figure*}
    \centering
    \includegraphics[scale = 0.5]{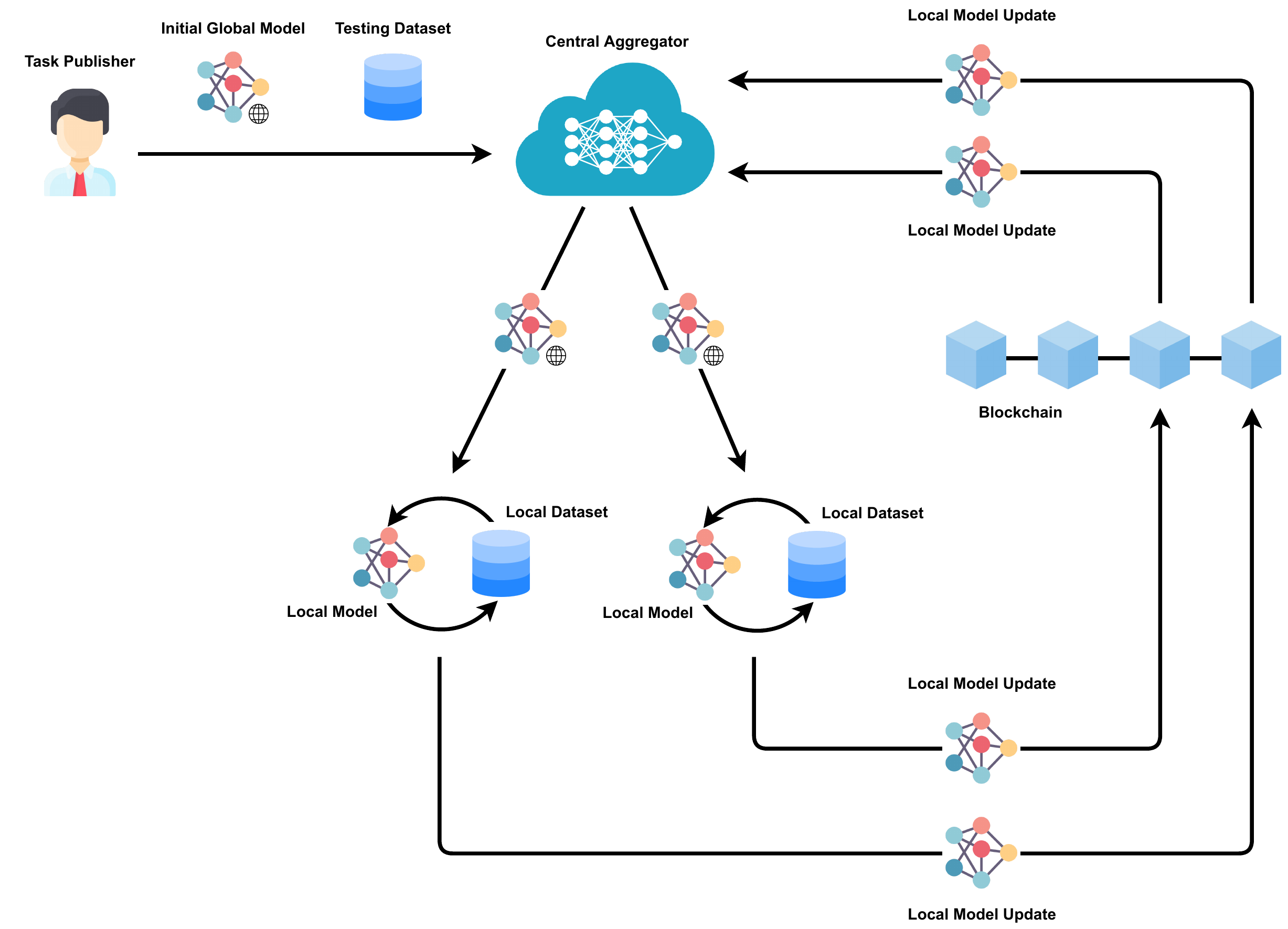}
    \caption[Short caption]{Typical Blockchain-Enabled Federated Learning Framework for Vehicular Networks}
    \label{fig:BC-enabled FL framework for IoVsS}
\end{figure*}

Figure \ref{fig:BC-enabled FL framework for IoVsS} illustrates a general implementation of Blockchain in FL for vehicular networks. A task publisher supplies a FL task, an initial global model, and a testing dataset to the blockchain system. Participant vehicles download the global model from the blockchain system and use it as a basis for training a local model using its own local dataset. After training its local model, a participant vehicle adds its local model update to the blockchain. Some frameworks such as \cite{pokhrel1} also utilize miners to collect local model updates and add them to Blockchain. Then either using a central aggregator that is a blockchain node or within the blockchain process, a node retrieves the local model update and aggregates all local models to form a new global model. After aggregation, the vehicle participants download the new global model and the process repeats for multiple iterations until the global model reaches a preset accuracy or optimal convergence.

\subsubsection {Blockchain-Based FL for Autonomous Vehicles}

In \cite{pokhrel1}, authors propose a Blockchain-based FL system for autonomous vehicles (AVs). The framework is composed of miners and AVs. Miners could be nodes on the network edges such as cellular base stations or AVs with high computing capabilities for the mining process. The authors use Blockchain in lieu of a central server to solve the centralization problem and develop a loyalty mechanism to reward AVs for providing good learning model updates. The reward amount is based on the usefulness of the data sample sizes. The framework also rewards miners who win the mining race. The mining reward is based in the miner's mining power, which is calculated by the number of useful data samples they collect from AVs. 

Under the authors framework, AV's obtain latest global model from Blockchain. Based on this model, AVs perform local learning and send a local model updates to a nearby miner. Since each AV maintains their best ML model, they are capable of making intelligent decisions even when they lose connectivity. All miners verify the local model updates, execute proof-of-work (PoW) and then generate a block to record the validated local model updates. The aggregation of the local models is added to this block as a new transaction. To address the additional delays associated with Blockchain management, the authors conducted an accumulated end-to-end delay analysis that took into consideration PoW, communication delays and computation delays. Authors used this analysis to minimize the overall delay of the system by dynamically adjusting block arrival rate.

\subsubsection{Blockchain-Based FL for Vehicular Networks Enhanced by Model Update Verification \& Local Differential Policy} \label{sec:BC-based FL for IoVss enhanced by model update verification & LDP}

In \cite{liu2}, the authors proposed a secure FL framework that takes advantage of the capabilities of Blockchain smart contracts to protect a 5G Network FL framework from poisoning attacks. The authors supplement their framework with local differential privacy (LDP) technologies that add Gaussian noise to a participants local model updates to protect a participant from membership inference attacks \cite{dwork1}\cite{bebensee1}. The authors in \cite{qi1} borrowed this approach to implement a privacy-preserving Blockchain-based FL for traffic flow prediction.

To develop their privacy-preserving Blockchain-based FL for traffic flow prediction, the authors in \cite{qi1} utilized consortium Blockchain in lieu of a central server in FL to protect against SPOF attacks. They also utilized a modified version of dBFT as a consensus algorithm. The dBFT consensus algorithm is similar to PBFT except that it allows for the flexible changing of miners rather than requiring a fixed miner group. The authors supplemented dBFT with mechanisms to protect against security attacks. For example, they added model update verification to ensure only qualified model updates from participant vehicles are aggregated by the aggregator. In model update verification, each miner RSU uses a testing dataset provided by the task publisher to judge whether a local model update is qualified for aggregation. Only qualified local updates with than a given threshold (set by the task publisher) are presented for aggregation. This process helps the system defend against poisoning attacks that would taint the global model during aggregation. Inspired by \cite{liu2}, the authors in \cite{qi1} also use LDP to protect the vehicle participants from membership inference attacks. When a participant vehicle uploads a model update, Gaussian noise is added to disturb the location information of the participant.

\subsubsection{Blockchain-Based Galaxy Federated Learning for Vehicular Networks}


In \cite{hu2020gfl}, the authors propose a Blockchain-based FL framework called Galaxy Federated Learning (GFL). GFL's storage system is composed of two parts: (1) Ethereum Blockchain and (2) InterPlanetary File System (IPFS). Ethereum is a decentralized open source Blockchain first conceived by Vitalik Buterin \cite{buterin1} with built-in smart contract capabilities \cite{tapscott1}. IPFS is a decentralized P2P file storage system based on Blockchain \cite{benet1}.

GFL utilizes an innovative algorithm designed by the authors called Ring decentralized FL algorithm (RDFL) \cite{hu2020gfl}. RDFL uses Ring-allreduce algorithm in distributed ML, consistent hash algorithm, and knowledge distillation. When training across multiple nodes, each node would normally have to wait for all other nodes to finish their training during each training iteration. The Ring-allreduce algorithm solves this communication efficiency problem by arranging all nodes in a circle or ring where each node receives data from its left neighbor and transmits data to its right neighbor \cite{gibiansky1}. During the scatter-reduce step, nodes exchange data so that every node on the ring receives a chunk of the final result. During the allgather step, the nodes exchange those chunks so that all nodes receive the final result. GFL's RDFL uses a customized version of this algorithm, which is described in more detail below. Knowledge distillation involves compressing a large model into a smaller, distilled model \cite{hinton2}. Distilling the dataset greatly reduces the amount of data that needs to be transmitted, improving overall communication efficiency and reducing communication pressure. GFL also uses an encryption mechanism to improve transmission efficiency and ensure the security of the transmitted information. GFL uses both asymmetric encryption and symmetric encryption to encrypt every piece of information in Blockchain.

RDFL makes full use of a networks bandwidth and improves GFL's security and robustness under malicious node attacks. Under RDFL, after each node completes local training, each untrusted node, according to the consistent hash algorithm, sends its model parameters to the first trusted node in a clockwise fashion. By using the consistent hash algorithm, RDFL transfers model parameters of untrusted nodes to different trusted nodes, which reduces communication pressure. Also, each trusted node uses Ring-allreduce algorithm (modified for RDFL) to pass model parameters to the next trusted node in a clockwise fashion. This process repeats until eventually all nodes have the model parameters of all nodes and all nodes reaches a stable state. Ring-allreduce helps maximize bandwidth utilization for model parameters between nodes. After all nodes reach a stable state, RDFL uses knowledge distillation to transfer knowledge (contained in the model parameters) to its locally trained model parameters.

Using IPFS also helps GFL reduce communication pressure since IPFS transmits a 46-byte IPFS hash of the model parameters in lieu of sending the actual model parameters. RDFL uses model parameter synchronization, which reduces communication congestion. RDFL also provides better bandwidth, which helps solve the communication efficiency problem. To solve the model security problem, RDFL reduces the influence of malicious nodes by not allowing untrusted nodes to access model parameters from other nodes. In addition, RDFL encrypts IPFS hash of model parameters to avoid exposure on Blockchain and protect information security. Lastly, GFL solves the aggregation performance problem by using knowledge distillation and dynamic polarization ratio to improve generalization of model while ensuring safety of the model. According to the authors, knowledge distillation has greater accuracy than FedAvg as an aggregation method for non-IID data.

\subsubsection{Hybrid Blockchain Framework with Asynchronous FL for Vehicular Networks}

In \cite{lu1}, authors propose a hybrid Blockchain framework with asynchronous FL that addresses challenges such as the computing efficiency, shared data reliability, delays associated with FL, and heterogeneity. Due to the heterogeneous communication and computing capabilities of vehicles participating in FL, there may be different learning rates for each and this slows down the learning iteration to the speed of the slowest participant vehicle. By making the system asynchronous, authors’ framework resolves this heterogeneity problem by not requiring the system to wait on the slowest participant.

To enhance data security, training efficiency, and accuracy, the authors design a hybrid Blockchain mechanism called PermiDAG. PermiDAG is composed of two hybrid components: (1) Main permissioned Blockchain and (2) local Directed Acyclic Graph (DAG). DAG is a directed graph data structure that uses topological order and has been used to improve the efficiency and scalability of traditional Blockchain such as in IoT \cite{popov1} and 5G beyond networks \cite{karlsson1}. DAG also uses a more efficient cumulative PoW in lieu of traditional PoW.

In PermiDAG, RSUs maintain main permissioned Blockchain while vehicles share data using DAG. Main permissioned Blockchain is responsible for global aggregation while local DAG is responsible for asynchronous local training. Under PermiDAG, vehicles train their local models based on the global model retrieved from main Blockchain. They also retrieve verified local models from other vehicles from local DAG. They execute local aggregation to improve their own local model and then add that model to local DAG. After several iterations of local training and aggregation, delegate RSUs collect the current local models from local DAG and perform global aggregation. The global model is then broadcasted to all delegates. All global models are collected into a block by a rotating delegate leader and the candidate block is appended to the permissioned Blockchain after verification. 

\subsubsection{Hierarchical Blockchain-Enabled FL for Knowledge Sharing in IoV}

In \cite{chai1}, authors propose a hierarchical Blockchain framework for IoV which utilizes their proposed hierarchical FL algorithm for knowledge sharing. Knowledge sharing \cite{zhang2} involves vehicles learning environmental data through ML methods and sharing that learned knowledge with each other \cite{chai1}. The authors design their framework to be hierarchical by performing consensus at Ground Chains and a Top Chain level. For each region, RSUs perform consensus for vehicle learning models. At the Top Chain level, base stations perform consensus for the models from each region. According to the authors, this multi-model approach is more feasible for knowledge sharing in large scale vehicular networks. Using a hierarchical approach, authors' framework can more easily manage vehicular networks composed of multiple regions and where each region has its own varying local characteristics.

The authors' work is primarily focused on addressing the two primary problems of implementing knowledge sharing in vehicular networks: (1) ensuring the security and reliability of knowledge shared during the sharing process (i.e., prevent poisoning attacks) and (2) privacy issues involved in storing aggregated knowledge in the central server (i.e., centralization issue and SPOF attacks). Authors framework also addresses loyalty challenge by utilizing a trading market process that stimulates the sharing of learning models.

To address security and reliability in knowledge sharing, the authors enhance FL with Blockchain. They propose a Proof-of-Knowledge (PoK) consensus mechanism that is light-weight and more efficient than traditional PoW. PoK combines machine learning (ML) with Blockchain consensus by replacing the hash puzzle of PoW with a ML process. Unlike PoW, which selects the fastest solver of the hash puzzle to propose a new block, PoK is designed to select the best solver that achieved the most accurate ML result. During Ground Chain consensus, RSUs verify the training accuracy of vehicles using a test set. To address the centralization issue and protect against SPOF attacks, the authors replace the central server of FL with Blockchain.

\subsection{Summary Remarks} \label{BC FL summary}

This section discussed a new line of work that combines AI and blockchain for IoV. Specifically, we discussed blockchain-based FL for IoV. We first presented an overview of FL, their challenges, and how blockchain can be used to solve these challenges. Then, we discussed recent studies that had build FL for BIoV, along with thier approaches and major differences. 
Table \ref{table:BC-enhanced FL applications for IoVss} outlines the various approaches to enhancing traditional FL for vehicular networks with Blockchain and how such approaches address issues such as poisoning, membership inference and SPOF attacks. It describes how each approach addresses issues related to loyalty and communication efficiency.

It should be noted that Blockchain-based FL for IoV is a recent trend that has gone out in the last two years. Thus, there exist several challenges with these techniques and several associated future direction. We will discuss those briefly in section \ref{challenges2}

\begin{table*}[!t]
    \centering
    \caption{Blockchain-Enhanced Federated Learning Applications for Vehicular Networks}
    \label{table:BC-enhanced FL applications for IoVss}
    \begin{tabular}{|l|l|l|l|l|l|l|}
    \hline
        \thead{} &
        \thead{Consensus} &
        \thead{\makecell{Poisoning\\ Attacks}} &
        \thead{\makecell{Membership \\Inference Attacks}} &
        \thead{\makecell{SPOF Attacks}} &
        \thead{Incentive Mechanism} &
        \thead{Communication Efficiency}\\
    \hhline{|=|=|=|=|=|=|=|}
        \cite{pokhrel1} & PoW & \xmark & \xmark & \makecell[l]{Use Blockchain\\in lieu of central\\server to protect \\against SPOF\\attacks} &  \makecell[l]{Loyalty mechanism\\that rewards AVs \\for providing good \\learning model\\updates. Reward\\amount based on \\usefulness of the\\ data \& sample sizes.} &
        \makecell[l]{To address delays associated\\with Blockchain management,\\authors conducted accumulated\\end-to-end delay analysis\\taking into account PoW,\\communication delays,\\and computation delays.\\Used this analysis to minimize\\overall delay of system by\\dynamically adjusting block\\arrival rate.}\\
    \hline
        \cite{qi1} & dBFT & \makecell[l]{ Model update\\verification\\ mechanism in\\consensus to\\ensure only\\qualified model \\updates are\\aggregated} & \makecell[l]{LDP adds\\Gaussian noise\\to participant\\local model\\updates to\\protect against\\membership\\inference attacks} & \makecell[l]{Use consortium\\Blockchain in lieu\\ of central server\\to protect against\\SPOF attacks} & \xmark & \xmark \\
    \hline
        \cite{hu2020gfl} & RDFL & \makecell[l]{RDFL reduces\\influence of\\malicious nodes\\by not allowing\\untrusted nodes\\to access model\\parameters from\\other nodes} & \makecell[l]{Use encryption\\ mechanism to\\ensure security\\of transmitted\\information} & \makecell[l]{Use Ethereum\\Blockchain in lieu\\of central server\\to protect against\\SPOF attacks} & \xmark & \makecell[l]{Encryption mechanism helps \\improve transmission efficiency.\\IPFS \& RDFL consistent hash\\algorithm help reduce\\communication pressure \\Ring-allreduce algorithm helps\\maximize bandwidth utilization\\for model parameters between\\nodes.}\\
    \hline
        \cite{lu1} & \makecell[l]{Hybrid:\\modified \\DPoS \& \\PermiDAG} & \xmark & \xmark & \makecell[l]{Use permissioned\\Blockchain in lieu\\of central server\\to protect against\\SPOF attacks} & \xmark & \makecell[l]{Asynchronous system resolves\\problem of having to wait on\\slowest participant.}\\
    \hline
        \cite{chai1} & PoK & \makecell[l]{During \\consensus,\\RSUs verify\\training \\accuracy\\of vehicles\\ using a test set} & \xmark & \makecell[l]{Use Blockchain\\in lieu of central\\server to protect\\against SPOF\\attacks} & \makecell[l]{Trading market\\process used to\\stimulate sharing} & \xmark\\
    \hline
    \end{tabular}
\end{table*}


\section{Challenges and Future Opportunities in BIoV} \label{sec:challenges future opportunities}

In this section, we discuss the challenges and future opportunities in BIoV. In Section \ref{challenges1}, key issues in BIoV are highlighted. In Section \ref{challenges2}, we review challenges and opportunities in integrating FL for BIoV.

\subsection{BIoV-specific Challenges and Opportunities}
\label{challenges1}

BIoV offers security and  mitigates data manipulations due to  data immutability brought by Blockchain. However, due to openness nature of the Blockchain, it cannot directly guarantee several issues such as security, trust, and privacy explained as follows.

 \textbf{Ensuring Trust among entities in IoVs.}  Trust and privacy is major issue in IoV as it is hard to believe any entity in IoVs due to the possibility of malicious behavior as well as to keep one’s private identity during communicating with other vehicles.  The limitations of the centralized trust management system can be solved through distributed approach using Blockchain but there still major concerns regarding the trust, privacy and scalability in operation among the participants in our vehicular network system. Authors in \cite{liu1} propose a conditional privacy preserving announcement scheme integrated with Blockchain to solve these problems where proof of work and PBFT are used to make the consensus to reduce the cost.  Another work by \cite{yang1}, that proposes to check the validity of the messages using Bayesian Interference model and each vehicle with the source message is given a rating by the message receiver. Based on the rating provided by vehicles, RSU calculate trust and make a block using those data and then RSUs add their blocks in the Blockchain where both PoW and PoS are used. However, privacy is not well addressed which mitigates vehicles from sharing their information.

\textbf{Security and Privacy in BIoV.} All the vehicles connected in IoV, share enormous amount of data for communication and making decision. For this, it is really important to provide security and privacy to these shared data which can be achieved through Blockchain. Due to openness of public Blockchain, it is hard to achieve this privacy. Vehicular social network (VSN) \cite{fan1} is also a part of IoV where a lot of small IoT devices are connected so it is important to make the encryption scheme for lightweight for reducing communication overhead \cite{chen3} as well as introduce better consensus algorithm for consortium Blockchain \cite{shen2}. Privacy can also be compromised when any of the transaction made though BTC, link certain identities to their pseudonyms that is known as linking problem. To solve the problems related to bitcoin another kind of cryptocurrency Altcoin is introduced. It can provide firm privacy with its advanced security features that need to be focused in future \cite{zaghloul1}. Recently, Intel software guard extension (SGX) has been integrated with the Blockchain based IoV to solve the existing limitations related to security and privacy. But the consensus algorithm used in SGX such as PoET \cite{intel1} suffer from stale of chip problem which eventually creates centralization of rights that need to be solved for better performance \cite{bao1}.

\textbf{Edge Computing for BIoV challenges and Opportunities.} Fog/edge computing is also required for computing the data in IoV where it is required to keep the data integrity which can be improved by a proper combination of Blockchain, AI and ML \cite{bhat1, singh2, wang2}. In IoV, dynamic access control is required in decentralized system, but it costs high energy consumption as well as at the same time it is needed to provide a secured and trusted environment. For solving this problem, authors \cite{cinque1}, propose to leverage Fog/Edge computing to ensure scalability with energy efficiency and use the edge nodes (RSU) as the mediator between the sensor nodes (vehicles or end devices) and edge computing. In this system, ledger entity (RSU) hosts the Blockchain and with the help of trusted node it will update the trust value for the to be trusted node (TBTN). This model has only focused on the trust for the entities not the messages they share as well as it also suffers from privacy problem and prone to false praise attacks.

\subsection{FL for BIoV Challenges and Opportunities}

\label{challenges2}

Despite the many benefits provided by Blockchain-enabled FL implementations for IoV, these implementations still suffer from several challenges~\cite{9712424}. Table \ref{table:BC-enhanced} summarizes these challenges and it is discussed as follows.


\textbf{Forking.} Forking  happens because of  network delays, a vehicle may not always have the most recent block and consequently create a different chain that branches or "forks" from the main one \cite{pokhrel1}. Forking reduces throughput since only a single chain survives. Forking
creates a situation where some vehicles  may utilize wrong global update in the next iteration of local model
update. The work on~\cite{pokhrel1} analyzed forking issues and its impact on FL on BIoV. Based on the analysis results, the  forking depends  on two factors;  block arrival rate and  block propagation delay. The work also developed a mathematical model have been developed to to capture the FL parameters such as  the retransmission limit,  block arrival rate, block size and the frame sizes so as to capture their impact on the system-level performance.

\textbf{Aggregation node selection}. The process of uploading and downloading model updates to FL could take a lot of time especially in complex models with many model parameters. Thus,  the selection of the  node responsible for aggregation and impact of that selection should been discussed. The work in~\cite{hu2020gfl} uses the amount of data as the indicator for the selection of aggregation nodes in FL.  However, other factors such as the number of computing resources of the data provider, current power, network conditions and other factors could also be used to improve the selection of aggregation nodes.   Future research work  could use methods such as model quantization and model distillation  to compress model parameters to improve FL speed and efficiency.

\textbf{Uncooperative Vehicle Training.} This challenge happens because of some vehicles are not willing to cooperate in the training phases in the FL.  Vehicles need to cooperate to build a high-quality global model and at the same time take into consideration communication resource allocation. The authors in \cite{wang3} address this issue in the realm of mobile devices and MEC. They propose a cooperative FL framework for the MEC system which they call CFLMEC, where devices transmit local models in a relay fashion with the goal of maximizing the admission of data at the edge server or nearby device. A similar framework could be incorporated into a Blockchain-based FL framework in vehicular networks to solve the uncooperative training strategy challenge. Another approach is exploring reputation and incentives systems to motivate all vehicles to to cooperate in the training phases in the FL. However, using reputation systems requires real identity of vehicles and this leads to another serious problem because vehicles will be reluctant to  cooperate in the training phases because of private information may be revealed because of adversarial attacks such as model inversion attacks~\cite{usynin2022beyond,chen2022inversion} and membership inference attacks~\cite{zhang2022membership,rastogi2022explaining}.

\begin{table}[!t]
      \centering
      \caption{Summary for FL challenges in  BIoV}
      \label{table:BC-enhanced}
\begin{tabular}{|c|l|}
\hline
     Challenge & Description \\\hline
    \hline
    \multicolumn{1}{|m{2cm}|}{Forking} & \multicolumn{1}{|m{5cm}|}{Vehicle fails to receive most recent block due to network delay and consequently creates a different chain that "forks" from the main one.} \\
    \hline

    \multicolumn{1}{|m{2cm}|}{Aggregation Node Selection} & \multicolumn{1}{|m{5cm}|}{Insufficient factors used for selecting nodes for aggregation.} \\
    \hline
        \multicolumn{1}{|m{2cm}|}{Uncooperative Vehicle Training} & \multicolumn{1}{|m{5cm}|}{Vehicles use an uncooperative training strategy.}\\
    \hline
         \multicolumn{1}{|m{2cm}|}{Communication and computation  Efficiency} & \multicolumn{1}{m{5cm}|}{Using iterative optimization techniques, which require many rounds of communication between local devices (e.g., vehicles) and the central serve.r}\\
         \hline
            \multicolumn{1}{|m{2cm}|}{\makecell[l]{Security \& \\Privacy}} & \multicolumn{1}{m{5cm}|}{Vehicles are reluctant to share models results because of cyberattacks such as eavesdropping  and spoofing attacks.}\\
  \hline
  \end{tabular}
\end{table}

\textbf{Communication and computation
Efficiency.} Current FL methods involve training models using iterative optimization techniques, which require many rounds of communication between local devices (e.g., vehicles) and the central server. This tends to lead to bottleneck issues and increased communication costs. To address this issue, the authors in \cite{guha1} proposed a one-shot FL approach, where the initial global model can be learned from data on the network in only one round of communication between devices and the central server. Zhou et al. \cite{zhou2} proposed distilled one-shot FL (DOSFL), which  combines the ideas behind data distillation \cite{wang5} and one-shot FL. Data distillation involves compressing a large dataset with thousands of images into a few synthetic training images\cite{wang5}. Realizing that using this compression method could help reduce communication cost in FL, Zhou et al. \cite{zhou2} apply distillation to one-shot FL. Under DOSFL, Each device distills their private dataset and uploads synthetic data to the central server instead of sending bulky ML gradients or weights. The central server then interweaves the distilled data together and uses it to train the global model. A similar approach could also be applied to Blockchain-based FL for vehicular networks.

\textbf{Security and Privacy of FL for BIoV.} Vehicles needs to exchange information during the FL training. Sharing this data shall be secure against different types of security and privacy-leakage attacks. The work in \cite{zhou2} designed their FL to be secure against eavesdropping attacks since an eavesdropper cannot reproduce the global model with leaked distilled data without knowing the exact initial weights distributed by the server. However, there may still exist security risks from other attacks such as data poisoning attacks. Future work could explore the potential attacks  resulted from using  Blockchain to secure FL frameworks against such attacks. Privacy techniques have been proposed to  aggregate the model updates, however, an untrusted aggregator can maliciously modify the aggregated result to produce a biased FL model.

\subsection{Performance Issues}

\label{perfromance}

Performance is major issue when using Blockchain to secure IoV. Performance can be due to several issues explained as follows.

\textbf{Scalability.} Blockchain  solves the limitations existed in centralized IoV system though it lags behind due to its poor performance when the number of vehicles and Blockchain nodes increases significantly \cite{salman1, lubin1}.   Factors that affect scalability includes but not limited to block size, number of transactions, total amount of connected nodes and consensus protocols used \cite{mazlan1}. Blocks with huge data packets cost much storage which leads a poor scalable system \cite{odonoghue1}. Due to high volume of data, it also results processing delay and communication delay which eventually degrade the performance of BIoV as most of the devices connected are resource constrained \cite{hussein1}. In BIoV, vehicles have to make transactions which need to be verified and synchronized such as in Ethereum it is required to validate each transaction part by part which causes slow speed and scalability challenges in the whole system \cite{mosakheil1, radanovic1, aljaroodi1}. Some works have discussed using zero-knowledge proves (ZKP) to solve security and privacy issues in BIoV. However, zero-knowledge proof (ZKP) can lead to computation intensity and Scalability issues due to its multiple round verification and limited throughput \cite{liang1, casado2021food,badr}. With the increasing of number connected vehicles in BIoV, latency is also incremented because of more computational requirement and affects throughput  in BIoV \cite{mcghin1, sanka1}. Thus, there are few recent protocols to address Scalability issues in BIoV but it lead to an increase in latency at the expense of high scalability \cite{agbo1}.

Developing efficient consensus algorithms tailored for BIoV is an open issue that can lead to improved Scalability. Delegated byzantine fault tolerance (dBFT) \cite{neo1} was first proposed by Da Hongfei and Erik Zhang for use in their NEO Blockchain cryptocurrency framework. The dBFT determines validators based on real-time Blockchain voting. This arrangement provides enhanced efficiency such as faster block time and transaction confirmation. Participant holders of a NEO token vote on a bookkeeper. The elected bookkeepers utilize BFT to reach consensus and generate new blocks.  dBFT has been employed by authors such as \cite{qi1} for Blockchain-based vehicular networks. Based on the discussion it is tangible that throughput and latency ate the two major factors that put a significant impact on the quality of experience (QoE) in BIoV that should be improved in future research \cite{chen2, zhou1}. The work in \cite{singh1} uses the shard Blockchain to reduce work loads in main Blockchain and smart contracts to make it more feasible. The main idea is that RSUs are responsible for maintaining and updating the trust of vehicles where proof of work consensus mechanism is used. Both permissioned and permission less Blockchain is integrated in this model for the main and shard Blockchain respectively. Though this model can improve scalability, it is not fully decentralized as some operations are controlled by Traffic Authority.

\textbf{Computations Overheads.} Performance of any BIoV depends on computation required. Encryption, decryption, and signatures that are required for the security purposes also require mining nodes with high computation. To utilize the vacant resources outside of Blockchain network such as cloud/fog computing by the evolutionary miners (E-miner) who are promoted from traditional miners (T-miner) with additional abilities for providing extra computing services.

During mining, each vehicle or RSU gets reward but in PoW based Blockchain networks sometimes miner acts as malicious node in incentive mechanism step which cause a waste of distributed computing resources. Zero-determinant strategy (ZD strategy) \cite{press1} is used to solve this challenge to encourage cooperative mining \cite{tang2}. Though it doesn’t take the trust value of the nodes into consideration which need to be solved in future for secured performance.

\textbf{Communication Overhead.} In IoT a lot of devices are connected, and they require frequent changes in the provenance data which causes enormous energy consumption that leads to computation loads, delays, and bandwidth overhead. With the increasing of the vehicles connected in IoV, energy needs are limited \cite{sekaran1}. That is why Blockchain network fails to provide good performance due to more energy consumption, execution time and communication overhead that need to be improved.

A selective compression scheme has been proposed to reduce communication overheads using a checkpoint-chain that can prevent the accumulation of compression. Enormous number of blocks can be verified with only a few updated checkpoints and nodes with limited resources can reduce the storage volume for the Blockchain ledger and achieve high verification capabilities \cite{kim1}.  A message passing technology called HPBC has been implemented in a high performance Blockchain consensus algorithm with byzantine nodes to solve the resource overhead and long block confirmation delay \cite{jiang2} but it does not consider privacy for the nodes as well as the malicious activities inside the nodes which should be studied to make its performance better.

\textbf{Designing Lightweight BIoV Protocols.} Security protocols include but not limited to  user authentication, key management, access control/user access control, and intrusion  prevention/detection. All these BIoV Protocols uses several cryptosystems that uses  heavy algorithms with
 high communications, computations and storage overheads. Some of the devices in  BIoV such as sensors are resource-constrained and lacks high communications computation, and storage capabilities~\cite{deebak2022lightweight,9713758}. Also, using Blockchain leads to extra overheads when it requires such capabilities. Thus,  using lightweight cryptosystems  can be helpful and more  research is needed to provide lightweight BIoV Protocols. Another issue is to develop efficient cryptosystems such as certificateless signcryption~\cite{xu2022certificateless}  which  can achieve data confidentiality and authentication without relying on a trusted third party to generate certificates. However, these cryptosystems comes with extra commutation and communication overheads and thus more research is needed to build lightweight schemes suitable for BIoV environments.

\section{Future Research Directions in  BIoV}
\label{sec:future-research-directions}
This section focuses on a number of future research directions in integrating Blockchain in AI, hardware security and Quantum computing.

\subsection{AI}

One future direction in the combination of AI and Blockchains is in the field of explainable and responsible AI. A current challenge in AI systems that hinder their adaptation in practice is that the decision made by these systems are untraceable and not explainable to humans. One solution to resolve this issue is to track the development of data flow and complex behaviors generated by the system. An example in IoV is the explaining why a model has stated that the traffic is congested. Such explanation can come from tracking data coming sequentially from different roads in the system. However, providing a traceable and transparent system that can provide such tracking without breaking user privacy is challenging. Blockchains can be used to perform such tasks by tracking every turn in the data-processing or decision-making chain. Such tracking can understand and increase confidence in AI decisions. Furthermore, tracking can be used by humans to trace back the decision process and justify decisions based on prior history or development. Lastly, tracking can provide insights into tuning AI systems to balance models\textquotesingle accuracy with their explainabilities. Of course, data and model privacy concerns would rise here, and so we do expect this direction to be directed toward applications that require only user privacy.


Another direction is to use AI to design lightweight BIoV protocols. As discussed in Section.~\ref{perfromance}, performance issues in BIoV is a major issue that need to be addressed to develop scalable and efficient BIoV. The intelligence from AI can help resolve these challenges by optimizing communication and computation overhead and providing the required scalable and efficient BIoV . 

In addition, while the integration of AI and Blockchains has many promises, it comes with many challenges and open issues that have been discussed in section \ref{challenges2}. Solving these challenges is a future direction to guarantee the success of Blockchain-based AI solutions. First and foremost, as Blockchains support transparency, the data stored in Blockchains is open and can be accessed by other nodes. This enforces AI data or model privacy concerns, especially if the data or model contain sensitive information. Note that the challenge here is data privacy, or model privacy in the case of FL, rather than user privacy. As discussed before, user privacy can be solved with the use of Blockchains. Solutions to data or model privacy can use private Blockchains to resolve this challenge. However, these solutions come with the limitations of providing heterogeneous environments, sometimes necessary for IoV. While the work in \cite{liu2} and \cite{qi1}  have proposed privacy-preserving Blockchain solutions, these solutions are still insufficient and at an early stage. Thus, a future direction should focus on Blockchain and AI integration while keeping the AI data or model private.

Another active research direction is the use of AI for Blockchains, i.e., building intelligent self-generated contracts, designing better AI-based mining techniques, and detecting malicious behavior and threats in Blockchains using AI techniques. This specific direction is a more general blockchain systems future direction and not specific to BIoV. However, we believe that such direction will revolutionize Blockchains and so it is important to be mentioned as a future direction.

\subsection{Hardware Security}

The IoV relies on embedded hardware, real-time sensors, ECUs, and printed circuit board (PCB), etc. to communicate critical information using the internet efficiently. The modern supply chain allows for the IoV embedded hardware to be designed, fabricated, assembled, and tested at different locations across the globe involving multiple untrusted parties, making the designed intellectual propriety (IP) inherently insecure. When the intellectual property (IP) of the IoV embedded hardware is not entirely secure and trusted, no security solution at a higher level of abstraction can protect the entire system.  In that case, an attacker may utilize a chip level venerability to access the and retrieve critical information, i.e., the owner's personal information, vehicle data, route map, etc., or harm/steal this information. Therefore, the protection of the IoV embedded hardware against the evolving hardware attacks is becoming crucial.

 At the hardware level, our future direction aims to cover novel security approaches  that integrate Physical Unalienable Functions (PUFs) and the state of the art topic in hardware-based security with the Blockchain technology for secure, trusted, and reliable BIoV embedded hardware. For example, recent research proposes Hardware-Assisted Blockchain (PUFchain) for
sustainable simultaneous device and data
security in the internet of everything (IoE)
 \cite{Mohanty2020}. The framework can be deployed in connected vehicles and autonomous networks to employee PUF-based hardware security with multiple roles in the BIoV, including enhanced low-cost authentication, lower latency, and lightweight low energy consumption. Further, the integration of hardware-based security will enable trust in IoV application at the hardware level and countermeasure against physical and embedded systems attacks, reverse engineering, hardware obfuscation, and  hardware counterfeit detection and prevention.

At the communication level, our future directions aim to integrate novel hardware security solutions and Blockchain technology to strengthen IoV Controller Area Network (CAN) security. Such secure solutions will ensure the authenticity and reliability of the messages sent over the IoV network and processes by  CAN-based bus architecture \cite{Cultice2020, Siddiqui2016}. In this regard, hardware-based security is helpful for key generation and message authentication, including obtaining a robust, unique, and reliable hardware-based shared key between ECUs, allowing all message paths to be securely encrypted. Additionally, hardware security enable obtaining true random number generation (TRNGs) used for robust a hash-based message authentication code (HMAC) system with a counter to help protect the BIoV system against replay attacks and message tampering attempts. 

A hardware-assisted security technique proposed as a delay-based plug-in-monitor for intrusion detection in IoV controller area network (CAN) topology \cite{Wang2018}. The proposed intrusion detection method prototype can be implemented efficiently on re-configurable platforms, Filed Programmable gate Arrays (FPGAs), and embedded micro-controllers that are part of the CAN bus shields. Such a hardware-based BIoV scheme protects the integrity of the messages on CAN buses leading to a further improve the security and safety of autonomous vehicles. This future direction will address the delay-based hardware-assisted security techniques against malicious hardware modifications, including remote hardware Trojan (RATs) and fault injection attacks, stressing the need to protect BIoV against such emerging hardware attacks.

\subsection{Quantum Computing Attacks}

Quantum computing is a field that is based on quantum theory which explains the nature and behavior of energy and matter on the quantum (atomic and subatomic) level. Quantum computers  get their unique power through the ability for bits to be in multiple states at one time, where  tasks  can be performed using a combination of 1's, 0's and both a 1 and 0 simultaneously ~\cite{de2021materials}.  The work in~\cite{khalifa2019quantum,sun2019towards}  discussed how quantum computers may attack proof-of-stake blockchains. The work has showed that PoW is safer than proof of stake because  hash functions are less vulnerable  quantum computers. The work has showed also attacks on PoS blockchains  using both Grover\textquotesingle s algorithm \cite{grover1} and Shor\textquotesingle s algorithm \cite{shor2}.  The authors demonstrated that quantum computing algorithms are capable of solving specific problems more efficiently than traditional computing algorithms. The work has recommended quantum resilient signature schemes to mitigate them. On the other hand, Blockchain uses one-way cryptographic hashing to secure transactions. However, quantum computing algorithms may one day be able to break blockchain's one-way cryptography. Thus, Quantum attacks has gained recently active attention from researcher~\cite{kiktenko2018quantum, sharma2021securing}, but BIoV network shall be well designed and be quantum-resistant to be secure against 51\% and byzantine attacks.

\section{Conclusion} \label{sec:conclusion}


This survey paper presents the recent advances, developments and state-of-the-art topics of BIoVs networks and their implementations for several life-critical applications, including crowdsourcing, energy trading, traffic congestion reduction, collision and accident avoidance, and infotainment and content cashing. This paper focuses on an in-depth review of federated learning (FL), future directions of blockchain-enabled hardware security and trust and next-generation blockchain-based quantum and post-quantum computing, emerging research areas that, to our knowledge, are not an integral part of any current survey work in the existing BIoV research domain. For that, Section \ref{sec:intro} introduces and defines the BIoV architecture and applications. Also, this section identifies the goals and motivations of our current survey paper. Additionally, the applications, security and privacy challenges of the existing BIoV schemes are summarized. The limitations of such existing work on BIoV applications related to the topic of this survey are discussed to depict the novelty and uniqueness of our survey article. Finally, the contributions of the current paper are listed in a precise manner in the introduction section. 

Section \ref{sec:background} provides the needed background information on BIoV architectures and applications. As shown in table \ref{table:existing-BC-IoVs-surveys}, our survey covers a wide range of applications related to blockchain-based IoV and surveys a broader range of emerging blockchain-based IoV articles. In particular, we provide additional detailed and in-depth discussions on blockchain implementations on many topics, including AI, machine learning and FL for BIoV. Lastly, our detailed discussions on challenges and future opportunities offers a good starting point for researchers to advance the study further. Section \ref{sec:BC-based applications in IoV} presents in-depth knowledge about the applications of blockchain-based IoVs such as energy trading, crowdsourcing, vehicle platooning, traffic congestion reduction, collision and accident avoidance and infotainment and content caching. Section \ref{sec:BC FL for IoVs} reviews the FL-based implementations for blockchain-based IoVs. This section discusses the challenges, i.e., security and privacy issues, for traditional FL and provides the emerging proposed research utilizing blockchain-enabled frameworks and schemes to overcome these challenges. Section \ref{sec:challenges future opportunities} presents the current challenges for BIoVs followed by a presentation of and the countermeasures against these challenges and future directions of BIoV  \ref{sec:future-research-directions}.


\bibliographystyle{IEEEtran}
\bibliography{references}

\begin{IEEEbiography}[{\includegraphics[width=1in,height=1.25in,clip,keepaspectratio]{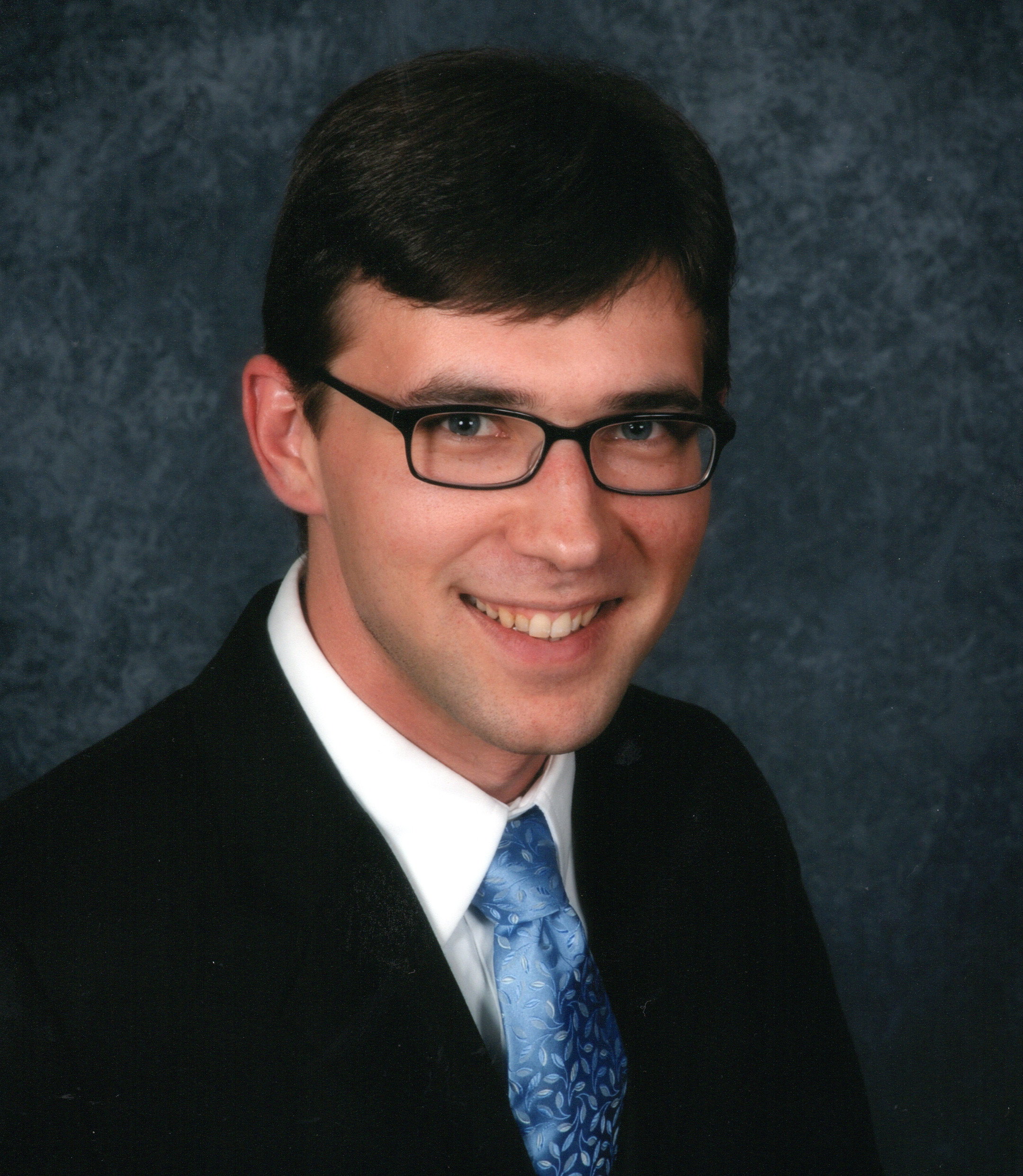}}]{\textbf{Brian Hildebrand}} is a PhD candidate in Technology with a focus on cybersecurity at GameAbove College of Engineering and Technology of Eastern Michigan University. He also serves as a Graduate Assistant and Instructor at Eastern Michigan University. He received a Master of Science in Computer Science from Eastern Michigan University in Dec. 2019 and a Juris Doctorate from Ave Maria School of Law in May 2010. For his Masters degree, Brian built an AI capable of composing music using a genetic algorithm and wrote a thesis on it. Brian coauthored a survey on blind signature for secure and trusted e-voting and e-cash systems that was featured in a chapter of \href{https://www.routledge.com/Security-and-Resilience-of-Cyber-Physical-Systems/Kumar-Behal-Bhandari-Bhatia/p/book/9781032028569}{Security and Resilience of Cyber Physical Systems}. His research interests include AI, machine learning, blockchain, IoV and cybersecurity.
\end{IEEEbiography}

\begin{IEEEbiography}[{\includegraphics[width=1in,height=1.25in,clip,keepaspectratio]{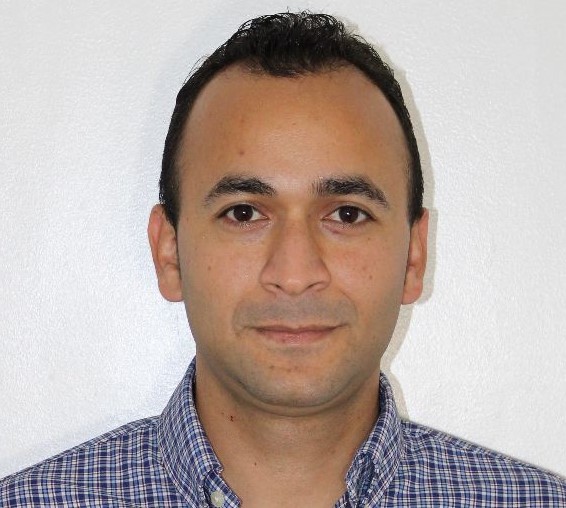}}]{\textbf{Mohamed Baza}} is currently an assistant professor at the Department of Computer Science at College of Charleston, SC, USA. He received his Ph.D. degree in Electrical and Computer Engineering from Tennessee Tech University, Cookeville, Tennessee, United States in Dec. 2020 and B.S. and M.S. degrees in Electrical \& Computer Engineering from Benha University, Egypt in 2012 and 2017 respectively. From August 2020 to May 2021, he worked as a visiting assistant professor at the Department of Computer Science at Sam Houston State University, TX, USA. He also has more than two years of industry experience in information security in Apache-khalda petroleum company, Egypt. Dr. Baza  is the author of numerous papers published in major IEEE conferences and journals, such as  IEEE Transactions on Dependable and Secure Computing, IEEE Transactions of Vehicular Technology (TVT), IEEE Transactions on Network Science and Engineering, and  IEEE Systems Journal, IEEE Wireless Communications and Networking Conference (IEEE WCNC) and IEEE International Conference on Communications (IEEE ICC). His research interests include blockchains, cyber-security, machine learning, smart-grid, and vehicular ad-hoc networks. He also a recipient of best IEEE paper award in the International Conference on Smart Applications, Communications and Networking (SmartNets 2020).
\end{IEEEbiography}

\begin{IEEEbiography}[{\includegraphics[scale=.3,clip,keepaspectratio]{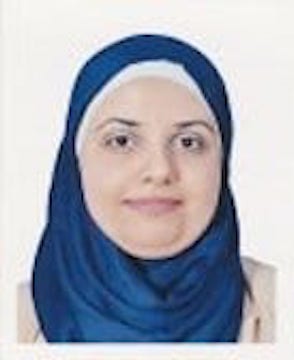}}]{Tara Salman} is an assistant professor of computer science at Texas Tech University, Lubbock, Texas, USA. Previously she was a Graduate Research Assistant at Washington University in St. Louis (2015-2021) and a Research Assistant at Qatar University (2012-2015). She received her Ph.D. from Washington University in St. Louis in May 2021 and her BS in computer engineering and MS degrees in computer networking from Qatar University Doha, Qatar in 2012 and 2015, respectively. Her research interests span Blockchains, network security, distributed systems, the Internet of things, and financial technology. She is an author of 1 book chapter, a patent, and more than twenty internationally recognized conferences and journals. 

\end{IEEEbiography}

\begin{IEEEbiography}[{\includegraphics[scale=.7,clip,keepaspectratio]{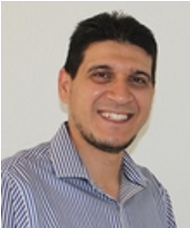}}]{Fathi Amsaad} received the bachelor's degree in Computer Science from the University of Benghazi, the top-ranked University in Benghazi, Libyan, in 2002, a dual master in Computer Science/Computer Engineering from the University of Bridgeport (UB), Bridgeport, CT, USA, in 2011/2012, and the Ph.D. degree in Engineering, with a Computer Science and Engineering concentration from the University of Toledo (UT), Toledo, Ohio, USA, in 2017. He is an Assistant Professor at Eastern Michigan University. His research is in "Hardware-Assisted Security for Smart and Embedded Systems". 
He author/co-authors 34 articles published in major IEEE conferences, journals, and Magazines in his research area, including IEEE International Workshop on Hardware-Oriented Security and Trust (HOST), IEEE World Forum on Internet of Things (WF-IoT), IEEE Transactions on VLSI, IEEE Open Access, and IEEE Consumer Electronics Magazine.  He served as the Technical Program Committee  Member for several IEEE conferences, including IEEE International Midwest Symposium on Circuits and System (MSCS) and IEEE International Symposium on Smart Electronic Systems (IEEE-iSES, formerly IEEE-iNIS). He serves as the Hardware/Software Vehicular Internment System (VIS) track chair in the 2021 IEEE-iSES. He is a reviewer and server in their review board for several journals and conferences, such as IEEE MSCS, IEEE ISCAS, Internet of Things Journal, IEEE Transactions on Computer-Aided Design, and MDPI Electronics Journal. He serves as Guest Editor in two MDPI-electronic Special Issues, including  "Special Issues: Hardware Intrinsic Security for Trusted Electronic Systems". He is an active IEEE member since 2011. 
\end{IEEEbiography}

\begin{IEEEbiography}[{\includegraphics[width=0.9in,height=1.15in,clip,keepaspectratio]{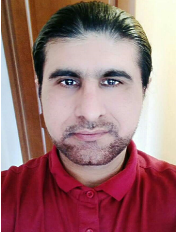}}]{Abdul Razaque} Abdul Razaque (Senior Member, IEEE) received the Ph.D. degree in computer science and engineering from the University of Bridgeport, USA, in 2015. He is currently a Professor with the Department of Computer Engineering and Telecommunications, International Information Technology University. His research interests include wireless sensor networks, cyber security, cloud computing security, design and development of mobile learning environments, and ambient intelligence. He has authored over 170 international academic publications, including journals, conferences, book chapters, and four books. He is an editor, an associate editor, and a member of editorial board of several journals.
\end{IEEEbiography}

\begin{IEEEbiography}[{\includegraphics[width=1in,height=1.25in,clip,keepaspectratio]{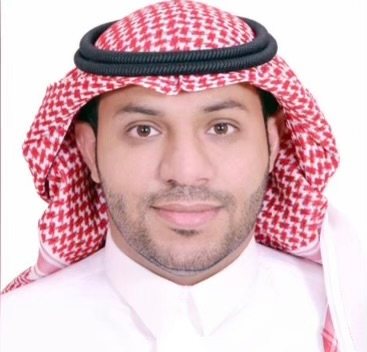}}]{Abdullah Alourani} is an Assistant Professor at the Department of Computer Science and Information, Majmaah University, Saudi Arabia. He received his Ph.D. in Computer Science from the University of Illinois at Chicago, his Master’s degree in Computer Science from DePaul University in Chicago, and his Bachelor’s degree in Computer Science from Qassim University, Saudi Arabia. His current research interests are in the areas of Software Engineering, Security, and Artificial Intelligence. He is a member of ACM and IEEE.
\end{IEEEbiography}

\end{document}